\documentclass[graybox]{svmult}
\pdfoutput=1
\usepackage{mathptmx}       % selects Times Roman as basic font
\usepackage{helvet}         % selects Helvetica as sans-serif font
\usepackage{courier}        % selects Courier as typewriter font
\usepackage{type1cm}        % activate if the above 3 fonts are
\usepackage{makeidx}         % allows index generation
\usepackage{graphicx}        % standard LaTeX graphics tool
\usepackage{multicol}        % used for the two-column index
\usepackage[bottom]{footmisc}% places footnotes at page bottom

\makeindex             % used for the subject index
\usepackage{amsmath} % AMS Math Package

\begin{document}

\title*{Fluctuations in active membranes}
\author{Herv\'{e} Turlier and Timo Betz}
\institute{Herv\'{e} Turlier \at Center for Interdisciplinary Research in Biology, Coll\`{e}ge de France, PSL Research University, CNRS UMR7241, Inserm U1050, 11 place Marcelin Berthelot, F-75005 Paris, France, \email{herve.turlier@college-de-france.fr}
\and Timo Betz \at Institute of Cell Biology, Center for Molecular Biology of Inflammation, Von-Esmarch-Str. 56, D-48149 Muenster, Germany, \email{timo.betz@uni-muenster.de}}

\maketitle

\abstract{Active contributions to fluctuations are a direct consequence of metabolic energy consumption in living cells. Such metabolic processes continuously create active forces, which deform the membrane to control motility, proliferation as well as homeostasis. Membrane fluctuations contain therefore valuable information on the nature of active forces, but classical analysis of membrane fluctuations has been primarily centered on purely thermal driving. This chapter provides an overview of relevant experimental and theoretical approaches to measure, analyze and model active membrane fluctuations. In the focus of the discussion remains the intrinsic problem that the sole fluctuation analysis may not be sufficient to separate active from thermal contributions, since the presence of activity may modify membrane mechanical properties themselves. By combining independent measurements of spontaneous fluctuations and mechanical response, it is possible to directly quantify time and energy-scales of the active contributions, allowing for a refinement of current theoretical descriptions of active membranes.}

\section{Introduction}
%To understand the term "`active membrane fluctuations"' it is helpful to quickly review the physical properties of biological and bio-mimetic phospholipid bilayer membranes, of long-standing interest in both biology and biophysics. 
Biological and bio-mimetic membranes consist in bilayers of phospholipids, which can embed various transmembrane or peripheral proteins. They constitute a selectively permeable barrier between distinct biological compartments, such as the cytosol and extracellular medium. The stability of lipid bilayers in water is the result of an entropic effect, which combines non-covalent interactions between hydrophobic and hydrophilic parts of lipids. This leads to a large in-plane rigidity, making lipid bilayers almost incompressible: area strains of only 2-4\% are generally enough to rupture a membrane. The non-covalent nature of interactions make phospholipid bilayers moreover tangentially fluid. Any tangential force on a lipid or embedded protein will lead to lateral flows balancing almost instantly any density gradient. Bending a lipid bilayer, in contrast, requires only to slightly displace the polar heads, which are separated by a distance of the order of 0.5-1 nm \cite{Campelo:2014, Hochmuth:1983}. The bending modulus, generally denoted $\kappa$, is therefore not very large compared to thermal energy, of the order of a few tens of $k_{\rm{B}}T$, which explains why membrane bending modes are readily excited at ambient temperatures. Hence, biological membranes are continuously fluctuating as a result of the thermal agitation of the surrounding medium, and this movement is directly observable using standard microscopy. However, besides thermal agitation, non-equilibrium active forces, of intrinsic or extrinsic origins, may also contribute and enhance membrane fluctuations. If such active fluctuations have random, uncorrelated sources, it remains however complicated to determine by simple observation to which extent membrane undulations are driven by thermal or by non-equilibrium effects. 

One of the most prominent example of fluctuating biological membranes is the 'flickering' of red blood cells, already described in the 19th century \cite{Browicz:1890}. Since its first observation, the origin of red blood cell flickering had been debated and only recently its possible active nature has been precisely investigated \cite{Tuvia:1997, Park:2010, Betz:2009, Turlier:2016}. Initially, flickering was suggested to be passive, similar to the Brownian motion of microscopic particles  \cite{Cabot_1901, Pulvertaft_1949}. But in 1951, the amplitude of flickering was shown to be correlated to ion transport across the membrane \cite{Blowers:1951}, suggesting a possible active metabolic driving. However, this metabolic interpretation was soon revised as flickering was also observed in the absence of ATP in red blood cell ghosts \cite{Parpart:1956}. After this finding, the pure passive origin of the flickering was generally accepted for more than 40 years \cite{Brochard:1975}, until in 1997 new experimental approaches revealed a change in the membrane fluctuations amplitude upon ATP starvation \cite{Tuvia:1997}. These new experiments, however, remained debated and a series of conflicting results were reported \cite{Evans:2008}. While most of the differences may be attributed  to variations in preparation protocols, more and more indirect findings suggested that an active driving may contribute to the low frequency fluctuations spectrum \cite{Park:2010, Rodriguez:2015}. A conclusive experimental evidence for the active nature of red blood cell flickering was recently given by comparing directly flickering and mechanical response of the membrane \cite{Turlier:2016}. The experimental observations could show that the flickering directly violates equilibrium statistical mechanics,
%at timescales slower than 100ms
proving the presence of non-equilibrium active forces driving membrane movement. This is an emblematic case of scientific controversy that took about 125 years to be conclusively resolved. 

As this example attests, it remains difficult to evaluate to which extent active processes may contribute to membrane fluctuations. In general, active fluctuations are superimposing upon passive thermal fluctuations, and from an experimental point of view, active and passive fluctuations may share similar characteristics. Hence, active contributions might not be visible, especially if thermal agitation dominates the fluctuation spectrum at the specific membrane position, length scale or timescale of interest. In this view, to answer the question whether fluctuations are active or passive, one should always define the relevant time and length scales involved.    

\begin{figure}
	\centering
		\includegraphics[width=1.0\textwidth]{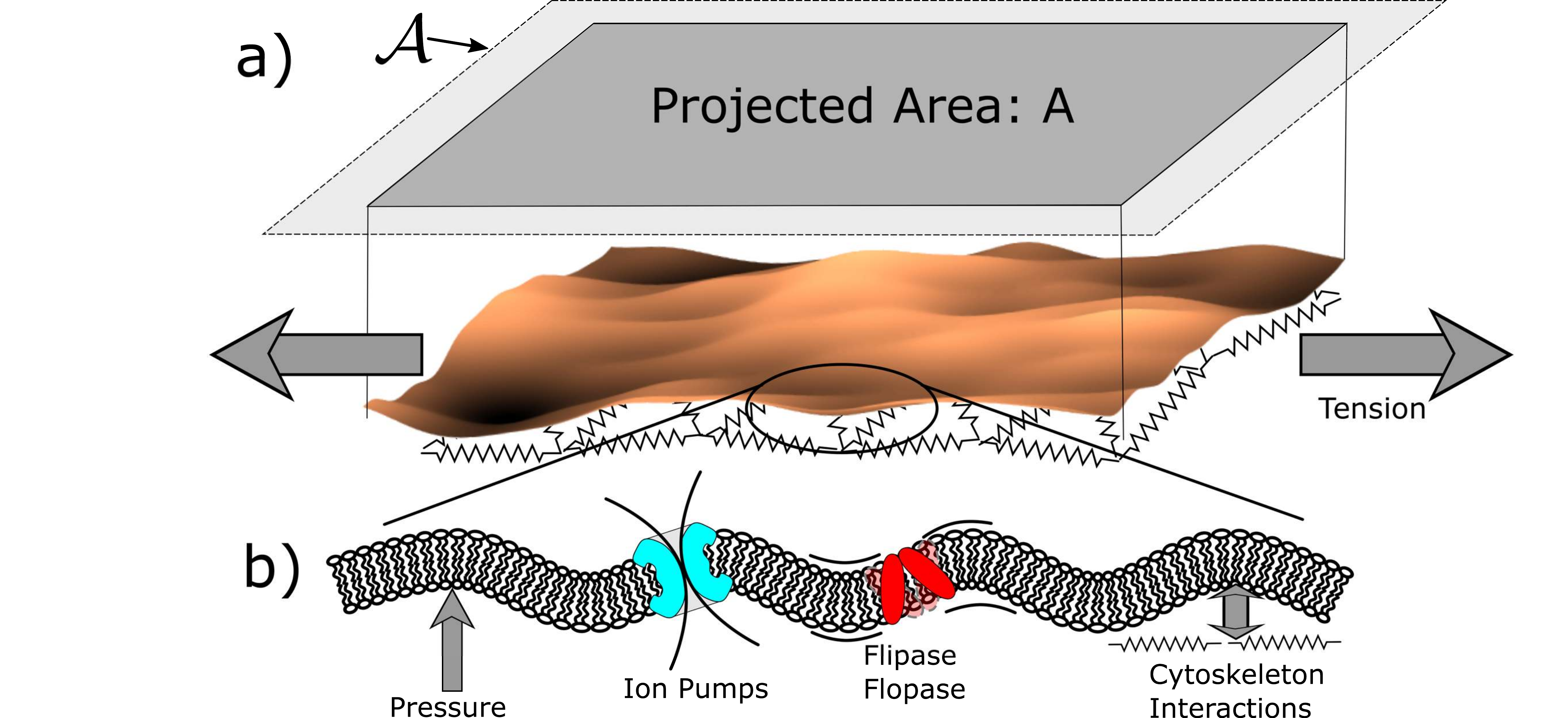}
	\caption{\textbf{Introduction of the concepts of membrane fluctuations, projected area surface tension and of the different active processes driving membrane fluctuations.} a) Thermal and active fluctuations buffer a significant fraction of the total membrane area, so that the projected area $A$ is smaller than the real area $\mathcal{A}$. Lateral pulling forces, effectively reduce the amplitudes of continuous fluctuations to extend the projected area. Hence these forces pull out the membrane reservoir that is stored in the fluctuations. The energy required to increase the projected area is used to define a membrane tension. It should be noted that this definition of tension depends on the entropic effect of thermally excited fluctuations. Biological membranes are typically connected to an underlying network of cytoskeletal elements such as F-actin or spectrin. b) Beside thermal agitation, membrane fluctuations can be driven by active, force generating processes such as ion pumps or lipid transporters activity or via mechanical coupling to the underlying cytoskeleton.}
	\label{fig:fig1}
\end{figure}

From a mechanistic point of view, active membrane fluctuations originate from the conversion of metabolic energy into forces by proteins inserted in the bilayer, or connected to it (Figure \ref{fig:fig1}a). One may define active membrane fluctuations as intrinsic, when they are produced by proteins directly embedded in the membrane, or extrinsic when activity originates from an independent structure tethered to the membrane, like the cytoskeleton.

Active membrane fluctuations originating from ion pumps have been the focus of pioneering biophysics studies over the last decades, both from theoretical \cite{Prost:1996, Ramaswamy:2000, Lin:2006} and experimental perspectives \cite{Manneville:1999, Manneville:2001, Faris:2009, Girard:2005}. These studies show that the activity of pumps leads to significant modifications in the fluctuation properties of reconstituted vesicles, measured as changes in the fluctuation amplitude, in the effective membrane tension or in the excess surface area. Besides ATP or photon driven ion pumps, lipid transport systems such as flipases and flopases \cite{Hankins:2015} may also contribute to active fluctuations, as well as membrane-fusion and fission of transport vesicles \cite{Rao:2001}. 

Active fluctuations can also originate from the interaction of the membrane with the underlying cytoskeleton, such as the spectrin network or the actomyosin cortex. In this case, the potential sources of active forces on the membrane are various: the proteins linking the membrane may change their binding affinity, or mechanical properties, upon phosphorylation, the cytoskeleton may exert tangential and normal forces on the bilayer under the action of molecular motors, or via polymerization of filaments.
\\

%The effect of active fluctuations on tension can be understood when analyzing the physical reason of the effective membrane tension in typical biological and biomimetic systems. The common definition of surface tension uses the energy required to change the surface area. In membranes, however, the real surface can only be modified by a few percent, as the membrane is almost incompressible and if stretched too much it ruptures. However, when observing the projected area or apparent surface (Figure 1a) it becomes evident that a significant amount of membrane area is hidden in the fluctuations. In other works the apparent surface is always smaller than the real surface of the membrane. When applying a tensile force to the membrane, the amplitude of the fluctuations is reduced and the apparent area increases.  This is at the base of the reported active membrane tension. As active forces now increase the amplitudes of the fluctuations, this effective tension is modified. 

The aim of this chapter is to present methods used to measure membrane fluctuations, to analyze their active and passive components, and to show how active membrane fluctuations may be modeled theoretically. 

\section{Experimental observations of active membrane fluctuations and indications of activity}
A challenging experimental task is to show that the observed fluctuations are active by nature. This requires separating the active and passive contributions in the fluctuation spectrum. A typical approach proposed to detect active fluctuations was to remove the source of chemical energy, primarily ATP, and then to attribute possible differences to active processes. Unfortunately, biological membranes are complex systems, and the suppression of metabolic energy sources does not only remove active noise, but it may also change the mechanical properties of the membrane. For example, ion pumps do not only contribute to active fluctuations but are also important for maintaining the osmotic pressure in the cell. This is, in turn, a key element determining membrane tension, and therefore influencing thermal fluctuations characteristics. It remains therefore generally challenging to disentangle fluctuations changes due to the suppression of activity from the ones due to passive mechanical variations. Another example is cytoskeletal softening or stiffening upon ATP depletion \cite{Humphrey_2002, Koenderink_2009}. In both cases, a simple removal of the energy source is not sufficient to pin down the contribution of active forces to the measured membrane fluctuations. In the following we will review different approaches that have been proposed to tackle this question in the context of membranes. 

	\subsection{Micropipette aspiration: surface tension and excess area}
	%Maneville, Evans, indirect measure of fluctuations as membrane storage
Surface tension is formally defined as the energy required to change surface area, but in lipid bilayers, the real surface can only be modified by a few percent before rupturing. However, a significant amount of membrane area is stored in the fluctuations, leading to clear difference between the total membrane area and its projected area (or apparent surface) (Figure \ref{fig:fig1}a). In other words, the apparent surface is always smaller than the real surface of the membrane, and this difference is measured as the membrane excess area. When applying a tensile force to a membrane, the amplitude of the fluctuations is reduced and the apparent area increases, corresponding to a decrease of excess area (Figure \ref{fig:fig1}b). For a vesicle or a cell, where the volume and total membrane area are supposed constant, the excess area is on the contrary fixed. In this case, active forces are expected to increase the fluctuations amplitude compared to pure thermal driving, leading to an increase of tension (see section 4.2.4).
 
\begin{figure}
	\centering
		\includegraphics[width=1.00\textwidth]{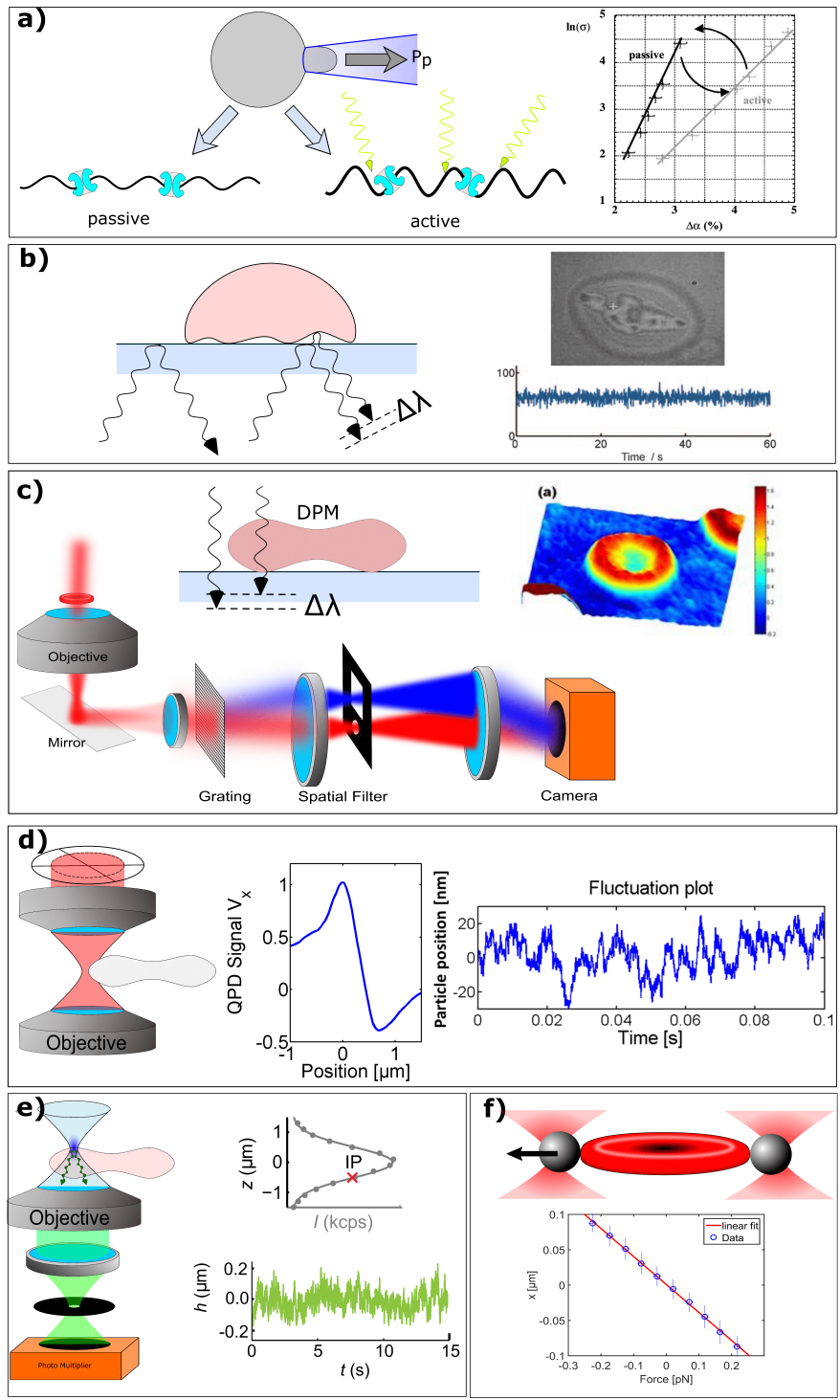}
	\caption{\textbf{Summary of the currently used techniques to study active and passive membrane fluctuations.} a) Active fluctuations can be measured as an increase of area stored in the membrane fluctuations upon illumination with light (Data reproduced  with permission from \cite{Manneville:2001}) b) A further method to determine membrane fluctuations is based on RICM, where constructive and destructive interference from reflections at the glass and membrane surface are detected with a camera.  c) More recently DPM has been introduce which measures thickness changes by exploiting the change in optical path along the light propagation through the object to be measured. (Inset adapted from \cite{Park:2006}, Figure 2) d) By exploiting the phase shift of light partially touching a membrane, the lateral fluctuations can be measured using an interferometric approach using a quadrant photodiode.  e) Another new technique related to fluorescence correlation spectroscopy can measure height fluctuations of a membrane with excellent spatial and temporal resolution. (Data reproduced from \cite{Monzel:2015}) f) Finally, direct membrane mechanics can be measured using optical tweezer based pulling on the membrane.  }
	\label{fig:fig2}
\end{figure}

Building on this physical reasoning, one of the first experiments showing that an active process can change membrane fluctuations was done in biomimetic liposomes, combining classical micropipette aspiration techniques with purified bacteriorhodopsin proteins. Bacteriorhodopsin is a light-driven transmembrane pump, which transfers protons across the membrane when exposed to green-yellow light of wavelength around 566 nm \cite{Wickstrand:2015}. Each time a proton is pumped, the membrane experiences a small active force. To measure this activity, micropipette aspiration was used. In this method the membrane area stored in fluctuations is measured by steadily increasing the aspiration pressure while following the changes in surface area. This change is calculated by following the length of the membrane tongue, while knowing the inner radius of the micropipette (Figure \ref{fig:fig2}a). In passive membranes, equilibrium statistical mechanics allows to predict that the logarithm of the applied tension is a linear function of the excess area. For the equilibrium case, the slope of this curve is proportional to $\kappa/k_{\rm{B}}T$ (see equation \eqref{eq:Surface_excess}). Intuitively a larger bending rigidity will indeed decrease the area stored in fluctuations, and a smaller excess area should hence be measured at the same stretching force. Conversely, increasing temperature will increase the area stored in fluctuations. When light sensitive ion pumps are activated by exposing the liposome to yellow-green light, the slope of the logarithm of tension vs. excess area curve is significantly changed, indicating a larger excess area stored in fluctuations. To explain the experimental result, the real temperature can be simply replaced by an increased effective temperature. Hence, the difference between the real and the effective temperature can be used to estimate the energy injected to drive active membrane fluctuations. In the case of bacteriorhodopsin the effective temperature was found to be about twice the real temperature \cite{Manneville:2001}. In a further experiment using a calcium pump driven by ATP hydrolysis, the effective temperature was found to be in the same range \cite{Girard:2005}. In these experiments a clear dependence of the effective temperature on the pump concentration was observed. 

Micropipette aspiration experiments provide hence a clear hint that membrane fluctuations are enhanced by active processes. Yet the actual fluctuations are only measured indirectly from the excess area. Other methods provide more direct access to membrane fluctuations and to possible active contributions. 

	\subsection{Image based contour analysis}
	%Sackman, Melard, Bassereau, pietro
Video microscopy provides direct spatial and temporal access to membrane fluctuations. Brochard and Lennon have pioneered its use to determine relative thickness fluctuations in red blood cells, responsible for the flickering effect \cite{Brochard:1975}.  At the time, an equilibrium model was used to analyze the membrane fluctuation spectrum, however as only relative amplitudes were measured it was not possible to extract mechanical properties from this model. This seminal work triggered a series of experiments aiming at inferring mechanical properties of biological and biomimetic membranes from precise measurement of their fluctuation spectrum. In a further improvement, image processing algorithms were developed to determine the time dependent fluctuation amplitudes as a function of lateral modes in liposomes and red blood cells. This was used by Sackmann and coworkers to gain experimental access to membrane properties such as the bending modulus and, in later work, to infer an effective tension value in red blood cells \cite{Zilker:1992, Strey:1995}. Over the following years, these techniques have been successfully refined, taking advantage of the rapid development in computer processing power and optical microscopy methods, paired with faster image acquisition methods \cite{Pecreaux:2004}. However, in all these works, pure passive membrane models were used to analyze the data, including the case of red blood cells. First approaches to study the fluctuation dynamics of active membranes were done again on the bacteriorhodopsin system, where a mode-dependent enhancement of the fluctuations was confirmed \cite{Faris:2009}. 
%The important advances were to finally directly observe the active fluctuations in driven systems, and to use an active model to describe these observations.
More recently, optical tweezers have been used to systematically excite well defined fluctuation modes and to subsequently study the mode-dependent relaxation behavior of the membrane \cite{Brown:2011}. Mode-dependent studies of active membranes allow to study possible complex mode-couplings due to activity or nonlinear interactions, which have remained overlooked so far. This approach may turn to be essential to study large biological membrane fluctuations, where the cytoskeleton may be the dominant source of active forces.
  
	\subsection{Interferometric methods: edge and height fluctuations}
	%Timo and Popescu/Park, RICM
Video microscopy analysis of membrane movement still suffers from limited spatial and temporal resolution and from complex and long image processing. To overcome this, a number of alternative methods have been developed that exploit interferometric approaches to gain sub-nm precision, with sometimes even $\mu$s time resolution. Here we will briefly discuss different approaches and their application to active membranes. Among the first interferometric techniques applied to measure membrane fluctuations is the Reflection Interference Contrast Microscopy (RICM, Figure \ref{fig:fig2}b) which allows to determine the distance between a membrane and the glass substrate with a resolution down to 1 nm \cite{Radler:1993}. Monochromatic light is reflected at the glass-medium interface as well as at the medium-membrane interface. Both reflected waves interfere on the camera either destructively or constructively, depending on the phase shift $\Delta \lambda$ of the light. The interference represents the distance between the membrane and the glass, and is used to detect height fluctuations of a membrane in close proximity to the glass. RICM has has been extensively used to study attachment phenomena both, in equilibrium and non-equilibrium situations \cite{Schmidt_2014, Monzel_2016}. 

A further interferometric method, that was used to detect active membrane fluctuations in red blood cells, is Diffraction Phase Microscopy (DPM, Figure \ref{fig:fig2}c). Here the optical path of the light traveling through the object creates an interference pattern where the zero and first order diffracted beams are arranged to interfere on a camera chip. The final image has diffraction limited resolution in the image plane, and nm precise resolution of membrane height fluctuations. Combined with a fast camera this allows spatially resolved membrane height fluctuations with a time resolution that essentially depends on the camera acquisition speed. Using this method it was recently shown that the positional probability distribution of the red blood cell membrane height has non-Gaussian contributions, that were proposed to result from active processes \cite{Park:2010}. However, also other possible explanation can give rise to such non-Gaussian behavior in a equilibrium situation, such as nonlinear force-displacement relations often found in biological systems. Interestingly, these additional contributions depend on the local curvature, with increased fluctuations at more curved regions. 

While RICM and DPM are sensitive to height fluctuations of passive and active membranes, another interferometric method was developed to determine the lateral fluctuations on a membrane at the rim of a cell or a liposome. The method is called time resolved membrane fluctuations spectroscopy and relies on a tight illumination of the membrane edge where part of the light penetrates the membrane and part is not interacting with the object (Figure \ref{fig:fig2}d). Typically this is done by a focused laser beam that is precisely positioned at the membrane interface \cite{Betz:2009}. As the common objects of interest provide a higher internal refractive index than the medium, the photons that traverse the object acquire a phase delay with respect to the photons that do not interact with the object. All the light is collected by a high numerical aperture condenser and the back focal plane of this condenser is imaged on a position sensitive detector or a quadrant photodetector. A calibration curve relates the position of the membrane to the signal on the detector and allows to determine the membrane fluctuations with sub-nm and $\mu$s temporal precision. The advantage of this method is that it is very fast, it does not require complex post-processing and it can be applied to any membrane oriented parallel to the axial direction of the laser focus. This technique was successfully used to determine the fluctuation spectra of red blood cells, growing membrane blebs and biomimetic liposomes \cite{Betz:2009, Betz:2012, Peukes:2014}. A similar technique derived from dark field microscopy was previously used to determine the relative fluctuations amplitudes of membranes in the context of red blood cells \cite{Tuvia:1997}. In the absence of a calibration procedure, this initial approach was not able to determine the absolute membrane position, but the relative measurement gave the first insights into possible active effects on red blood cell membrane fluctuations.  

	\subsection{Fluorescent detection of axial fluctuations}
	%Relation to FCS
Recently, a fluorescence based method has been used to determine height fluctuations of red blood cells and other cell types by combining a confocal imaging method with a fast detector. Called Dynamic Optical Displacement Spectroscopy (DODS) this technique is in principle closely related to fluorescence correlation spectroscopy, with the difference that not the lateral diffusion of membrane bound molecules, but the out-of-plane movement of the full membrane is measured. When a high concentration of fluorescent molecules is present in the membrane, it is not the number of molecules in the focus that dominates the signal, but the position of the membrane with respect to the focal plane. Since the detector is placed behind a confocal pinhole, only the fluorescent signal originating from the focal region is detected. The resulting fluorescence intensity as a function of the axial (z) position corresponds to the point spread function of the pinhole. Hence, the fluorescent intensity can be translated into the axial position of the membrane. This results in a axial precision below 20 nm and a temporal resolution of 20 $\mu$s. DODS successfully verified the non-Gaussian behavior of the membrane fluctuations in red blood cells and the dependence of these fluctuation on the local curvature of the membrane \cite{Monzel:2015}. In contrast to the DPM method introduced above, DODS has a superior time resolution similar to the time resolved membrane fluctuation spectroscopy. Furthermore, it is not limited to samples providing a homogeneous optical density, but can, in principle, be used on any object. 

	\subsection{Optical tweezers}
A direct method to probe active membranes is based on optical tweezers (Figure \ref{fig:fig2}f). Here the momentum transfer of a highly focused laser generates a net force on arbitrary shaped particles if their refractive index is higher than the surrounding medium. In a typical situation, micrometer sized beads are trapped, calibrated and then used to apply well defined forces on the object of interest. Initially, optical tweezers were used to measure the mechanical properties of red blood cells in the linear and nonlinear regimes \cite{Henon:1999, Mills:2004}. Many optical tweezer setups are also capable to detect the movement of single particles with very high precision. Here, the same methodology as for the time resolved membrane fluctuation spectroscopy is used, where a position sensitive detector measures asymmetries in the light deflected off the object in the beam path. The combination of these capabilities was used to determine both the mechanical response function of a red blood cell membrane, and the spontaneous fluctuation spectrum of the same cell. For this, polystyrene beads are attached to the membrane and serve simultaneously as handles to fix the cell in space, as probe particles that can apply precise forces and as position sensors to follow the membrane deformation. In a first experimental system two beads were attached to a red blood cell, and both the mechanical properties as well as the fluctuations were studied. However, in this approach the fluctuations were not fully free as the beads needed to be still optically trapped, and hence the free membrane fluctuations were partially suppressed by the trapping potential created by the laser \cite{Yoon:2011}. To explore the full fluctuation spectrum, a more complex four beads system was developed \cite{Turlier:2016}. Here, four beads are attached on opposing sides on the rim of the red blood cell, and three of them are trapped by a tweezer and serve as handles to maintain the cell at a well-defined and stable position. The fourth bead is either forced to oscillate with a well-defined frequency, or simply allowed to move freely without any force to measure the free membrane fluctuations. In this setup the probe bead is not restricted in its movement by a laser trap. If the membrane fluctuations are purely thermal, equilibrium statistical mechanics connects the membrane fluctuation characteristics with the dissipative response of the membrane, and the tweezer setup was used to check this correspondence. The power of this approach is that it does not only provide clean experimental access to systematically study the active fluctuations and mechanical properties of a membrane, but it also results in precise frequency-dependent data. This permits comparison to detailed theoretical models, which include temporal characterization of the active membrane fluctuations. 

\section{Analysis of active membrane fluctuation data} 
The different experimental methods described above provide various determination of membrane fluctuations, ranging from their indirect assessment by the excess area, to direct spatio-temporal measurements using video-microscopy. To give sense to these experimental data, analysis methods have been developed. They can be separated into static analysis, that presents typically time averages or histograms without consideration of time-dependent aspects, and on the other hand dynamic analysis concepts such as the auto-correlation function and the power spectral densities, which directly quantify temporal variations in membrane position. In general, dynamic analysis provides more extensive insight in fluctuations properties, but it requires stringent time resolution and a large number of data-points to produce statistically relevant results. 
%A special case of dynamic data analysis is presented by the frequency dependent measurement of the response function, which is discussed here separately.

Analysis of fluctuations requires to introduce an adapted theoretical formalism for the interpretation of quantitative data. The Helfrich's physical framework has proven its broad relevance to biological membranes since its introduction in 1973 \cite{Helfrich:1973}, and is briefly overviewed in the following for a membrane at equilibrium.

\subsection{Theoretical framework for equilibrium membrane fluctuations}

	\subsubsection{Helfrich's membrane Hamiltonian}
Helfrich proposed a surface energy density of the form $\frac{\kappa}{2}\left(2C-C_0\right)^2$ for the bending energy of a lipid bilayer, where $C$ is the local mean curvature and $C_0$ a spontaneous curvature (that we will generally ignore in the following). Note that we generally disregard an additional Gaussian curvature term, which reduces to a constant when integrated over a closed, or periodic surface. Since stretching lipids from one-another represents a high energy cost at the bilayer level, one may generally consider the membrane total area constant. From a theoretical point of view, this area constraint is enforced via a Lagrange multiplier denoted $\sigma$, and which is interpreted physically as the membrane surface tension. The Helfrich energy $\mathcal{H}$ is written as the sum of these bending and tension contributions, integrated over the total membrane area $\mathcal{A}$

\begin{equation}
	\label{eq:HelfrichHamiltonian}
	\mathcal{H} =  \int_{\mathcal{A}}\left\lbrace \frac{\kappa}{2}(2C-C_0)^2 + \sigma \right\rbrace d\mathcal{A}
\end{equation}

		\subsubsection{Static membrane fluctuations spectrum}
	We consider here a quasi-planar membrane of bending modulus $\kappa$ and surface tension $\sigma$, as sketched in Figure \ref{fig:fig1}a and we describe its shape in the Monge representation, where the position vector of the bilayer mid-plane is measured by the height vector $\big((x,y),\,h(x,y)\big) = \big(\mathbf{r},\,h(\mathbf{r})\big)$. In the limit of small deformations $\big| \boldsymbol{\nabla}h(\mathbf{r})\big| \ll 1$ the Helfrich Hamiltonian \eqref{eq:HelfrichHamiltonian} can be written, to a constant
\begin{equation}
	\label{eq:HelfrichHamiltonian_MongeRepresentation}
	\mathcal{H} = \int_A d^2\mathbf{r}\left\lbrace\frac{\kappa}{2}\big[\boldsymbol{\nabla}^2{h}(\mathbf{r})\big]^2 + \frac{\sigma}{2}\big[\boldsymbol{\nabla}h(\mathbf{r})\big]^2\right\rbrace
	\,\textrm{,}
\end{equation}
where $A$ denotes the projection of the membrane area $\mathcal{A}$ onto the plane $(x,y)$.

			\paragraph{\textbf{Static fluctuation modes}}
		To study fluctuations, it is more convenient to work in spatial Fourier space, and we consider hence a square piece of membrane with periodic boundary conditions. The membrane height in real space can be decomposed as $h(\mathbf{r}) = \frac{1}{A}\sum_{\mathbf{q}} h_{\mathbf{q}}\,e^{i\mathbf{q}\cdot\mathbf{r}}$ where $h_\mathbf{q} \equiv \int_Ad\mathbf{r}\,h(\mathbf{r})\,e^{-i\mathbf{q}\cdot\mathbf{r}}$ is the Fourier component for the wave-vector $\mathbf{q}$ of membrane deformation. Inserting this expression in the Helfrich Hamiltonian \eqref{eq:HelfrichHamiltonian_MongeRepresentation} yields
\begin{equation}
	\label{eq:HelfrichHamiltonian_FourierSpace}
	\mathcal{H} = \frac{1}{2}\sum_{\mathbf{q}}\left(\sigma \mathbf{q}^2 + \kappa \mathbf{q}^4\right)|h_{\mathbf{q}}|^2
	\,\textrm{.}
\end{equation}
	  
	  If the membrane is at thermal equilibrium, we can use the equipartition theorem, which assigns an average energy $k_{\rm{B}}T/2$ energy to each Fourier mode, yielding the static membrane fluctuation spectrum
\begin{equation}
	\label{eq:FluctuationSpectrum_Static}
	\left\langle |h_{\mathbf{q}}|^2\right\rangle = \frac{k_{\rm{B}}T}{\sigma \mathbf{q}^2 + \kappa \mathbf{q}^4}
	\,\textrm{,}
\end{equation}
where $\left\langle.\right\rangle$ denotes a statistical average.

The modes of deformation are limited for a real membrane, which has a typical size $L \sim \sqrt{A}$ and is composed of lipids of microscopic size $a\sim 0.5\,nm$. These two length scales allow us to define the following macroscopic and microscopic modes cutoffs: $q_{\rm{min}}\equiv 2\pi/L$ and $q_{\rm{max}} \equiv 2\pi/a$. In the reasonable limit $a \ll L$, the fluctuation spectrum amplitude can be calculated by integrating the static fluctuation spectrum \eqref{eq:FluctuationSpectrum_Static} over all wave-vectors in the range $\big[q_{\rm{min}}, \,q_{\rm{max}}\big]$, which yields
\begin{equation}
\left\langle |h|^2 \right\rangle = \frac{k_{\rm{B}}T}{4\pi\sigma}\ln\left(1+\frac{\sigma}{\kappa q_{min}^2}\right)
\label{eq:static_fluct_spectrum_integrated}
\end{equation}

From the equation \eqref{eq:static_fluct_spectrum_integrated}, one readily observes that a characteristic length $\lambda_{\sigma} = \sqrt{\kappa/\sigma} = 2\pi\,q_{\sigma}^{-1}$ may be defined from the bending modulus and surface tension. Two membrane fluctuation regimes can be identified, depending on the amplitude of membrane wave-vector $q = |\mathbf{q}|$ relative to the intrinsic length scale $q_{\sigma}$:

	\begin{itemize}
	\item If $q \gg q_{\sigma}$, the fluctuations are controlled primarily by modes in $q^4$, corresponding to the membrane bending elasticity term. In this regime, membrane fluctuations are dominated by the longest deformation wavelength, and their squared amplitude scales as $\left\langle |h|^2 \right\rangle \sim \frac{k_{\rm{B}}T}{4\pi\kappa}\frac{A}{\pi^2}$. 
	\\
	For a bilayer of typical size $L \sim 10\,\mu m$, bending modulus $\kappa \sim 10\,k_{\rm{B}}T$ and vanishing tension, one would obtain $\sqrt{\left\langle h^2(L) \right\rangle}\approx 3\,\mu m$. 
	\\
	\item If $q \ll q_{\sigma}$, the tension modes in $q^2$ dominate fluctuations, and one can evaluate the fluctuation squared amplitude as $\left\langle |h|^2 \right\rangle \sim \frac{k_{\rm{B}}T}{4\pi\sigma}\ln\left(\frac{\sigma}{\kappa\,q_{\rm{min}}^2}\right)$. 
	\\
	For typical lipid size $a\sim 0.5\,nm$ and membrane tension $\sigma \approx 10^{-5}\rm{N.m}^{-1}$, one obtains an amplitude $\sqrt{\left\langle h^2(L) \right\rangle}\approx 60\,nm $.
	\end{itemize}	

			\paragraph{\textbf{Excess area and membrane tension}}
We see above that the amplitude of thermal fluctuations is strongly reduced when the membrane is tensed. 
As we mentioned before, the membrane tension is in fact intrinsically related to the constraint of conserved membrane area. To formalize this relation one may introduce the area excess $\alpha$, which measures the difference between projected area $A$ and total membrane area $\mathcal{A}$ \cite{Helfrich:1984} (see Figure \ref{fig:fig1}a)
\begin{equation}
	\label{eq:Surface_excess}
	\alpha = \frac{\mathcal{A}-A}{\mathcal{A}} \sim \frac{1}{2A}\int_A d\mathbf{r} \,\left|\boldsymbol{\nabla}h(\mathbf{r})\right|^2 = \frac{k_{\rm{B}}T}{8\pi\kappa}\ln\left(\frac{q_{\rm{max}}^2 + \sigma/\kappa}{q_{\rm{min}}^2+ \sigma/\kappa}\right)
\end{equation}
For vanishing tension, this formula yields simply $\alpha = \frac{k_{\rm{B}}T}{4\pi\kappa}\ln\left(\frac{L}{a}\right)$.
For intermediate tension values $q_{\rm{min}} \ll q_{\sigma} \ll q_{\rm{max}}$, tension is directly related to the excess area as $\alpha = \frac{k_{\rm{B}}T}{8\pi\kappa}\ln\left(\frac{\kappa\,q_{\rm{min}}^2}{\sigma}\right)$. Knowing the membrane excess area, one can therefore calculate the membrane tension directly by inverting this relation.

Note that for higher tension values one needs to consider an additional term $\frac{\sigma}{K_c}$ to the excess area to take into account of the lipids stretching, characterized by a bulk modulus $K_c$ \cite{Fournier:2001}.

		\subsubsection{Dynamic fluctuation spectrum}
			\label{sec:DynamicFluctuationSpectrum}
			
			\paragraph{\textbf{Membrane Langevin dynamics}}
Biological or bio-mimetic membranes are embedded in aqueous solutions, which need to be displaced when the membrane deforms. Since inertia is negligible at this scale, fluid flow can be described by Stokes equations
\begin{eqnarray}
	 - \boldsymbol{\nabla} p(\mathbf{r}) + \eta \boldsymbol{\nabla}^2\mathbf{v}(\mathbf{r}) &=& - \mathbf{f}(\mathbf{r}) \\
	 \boldsymbol{\nabla} \cdot\mathbf{v} &=& 0
\end{eqnarray}
where $\mathbf{v}$, $p$ and $\eta$ are respectively the fluid velocity, pressure and viscosity, and $\mathbf{f}$ is a bulk force in the fluid.

The viscous force opposing membrane displacement can be inferred by calculating the flow generated by a point-like force $\mathbf{f}(\mathbf{r}) =  \mathbf{F}\delta(\mathbf{r})$ and integrating this force along the membrane. The Stokes flow solution is related to the point-like force $\mathbf{f}$ through a Green's function $\Lambda(\mathbf{r})$, called the Oseen tensor \cite{Doi:1988} and defined in real space as
$\mathbf{v}(\mathbf{r}) = \int \Lambda(\mathbf{r}-\mathbf{r}')\mathbf{f}(\mathbf{r}')d^3\mathbf{r}'$.
Following the derivation proposed in \cite{Doi:1988}, the diagonal part of the Oseen tensor may be calculated as
\begin{equation}
\Lambda(\mathbf{r}) = \frac{1}{8\pi\eta|\mathbf{r}|}
\end{equation}

To obtain the membrane dynamics, we note that normal velocities of the fluid and of the membrane should coincide at the membrane surface $\mathbf{v} = \frac{\partial h}{\partial t}$, and that the membrane exerts an instantaneous elastic restoring force per area on the fluid $f^{\rm{el}}(\mathbf{r},t) = - \frac{\delta\mathcal{H}}{\delta h(\mathbf{r},t)}$. By adding an additional white noise term in the force, to account for thermal agitation, we obtain an overdamped Langevin dynamics for the membrane height $h(\mathbf{r},t)$
\begin{equation}
\frac{\partial h}{\partial t}(\mathbf{r},t) = \int d^3\mathbf{r}' \,\Lambda(\mathbf{r} - \mathbf{r}') \left\lbrace - \frac{\delta\mathcal{H}}{\delta h(\mathbf{r}',t)} + \zeta^{\rm{th}}(\mathbf{r}',t)\right\rbrace
\end{equation}

In spatial Fourier space, this equation reads more concisely
\begin{equation}
\frac{\partial h_{\mathbf{q}}(t)}{\partial t} = \Lambda_{\mathbf{q}}\left[ - \left(\kappa \mathbf{q}^4 +\sigma \mathbf{q}^2 \right)h_{\mathbf{q}}(t) + \zeta^{\rm{th}}_{\mathbf{q}}(t)\right]
\label{eq:FourierLangevinEqDynamics}
\end{equation}
The thermal noise term has zero mean and its correlations obey the fluctuation-dissipation relation
\begin{equation}
\label{eq:EquilibriumNoiseCorrelation}
\left\langle \zeta^{\rm{th}}_{\mathbf{q}}(t)\zeta^{\rm{th}}_{\mathbf{q'}}(t') \right\rangle = 2k_{\rm{B}}T\Lambda_{\mathbf{q}}^{-1}\delta(\mathbf{q}+\mathbf{q}')\delta(t-t')
\end{equation}
where the Oseen tensor in Fourier space is given by $\Lambda_{\mathbf{q}} = \Lambda_q = 1/4\eta q$, with $q \equiv |\mathbf{q}|$.

			\paragraph{\textbf{Power spectral density}}
			\label{sec:PowerSpectralDensity}
From the membrane dynamics \eqref{eq:FourierLangevinEqDynamics}, we can directly compute in temporal Fourier space the mode-dependent autocorrelation function for an equilibrium membrane
\begin{equation}
	\left\langle \left| h_{\mathbf{q}}(\omega)\right| ^2 \right\rangle = \frac{2k_{\rm{B}}T\Lambda_{q}}{\omega^2 + \omega_{q}^2}
	\label{eq:ModeDependent_CorrelationFunction}
\end{equation}
where $\omega_{q}$ is the typical membrane relaxation rate for the mode $q$, given by
\begin{equation}
 \omega_{q} \equiv \Lambda_q\left(\kappa q^4 + \sigma q^2\right) = \frac{\kappa q^3 + \sigma q}{4\eta}
 \label{eq:MembraneRelaxationRate}
\end{equation}

To obtain the equilibrium membrane fluctuation spectrum (or power spectral density, noted PSD), one has to integrate the autocorrelation function \eqref{eq:ModeDependent_CorrelationFunction} over all deformation modes explored by the membrane
\begin{equation}
\left\langle \left| h(\omega)\right| ^2 \right\rangle = \int \frac{d\mathbf{q}}{(2\pi)^2} \,\frac{2k_{\rm{B}}T\Lambda_{q}}{\omega^2 + \omega_{q}^2} = \frac{4\eta k_{\rm{B}}T}{\pi}\int_{q_{\rm{min}}}^{q_{\rm{max}}} \frac{dq}{(4\eta\omega)^2 + (\kappa q^3 + \sigma q)^2}
\end{equation}

Supposing $q_{\rm{min}}\sim 0$ and $q_{\rm{max}}\sim \infty$, the PSD scales as $\omega^{-5/3}$ and $\omega^{-1}$ in the limit cases of, respectively, high and low frequency
\begin{subequations}
	\label{eq:PSDLimitScalingBehavior}
	\begin{equation}	
	\left\langle\left| h(\omega)\right| ^2 \right\rangle \underset{\omega\rightarrow\infty}{\longrightarrow} \frac{k_{\rm{B}}T}{6\pi(2\eta^2\kappa)^{1/3}\omega^{5/3}} 
	\end{equation}
		\begin{equation}	
	\left\langle\left| h(\omega)\right| ^2 \right\rangle \underset{\omega\rightarrow 0}{\longrightarrow} \frac{k_{\rm{B}}T}{2\sigma\omega}
	\end{equation}
\end{subequations}

			\paragraph{\textbf{Fluctuation-dissipation relation}}
			\label{sec:FluctuationDissipation}
To write the fluctuation-dissipation relation, we have to determine the mechanical response function of the membrane, defined in temporal Fourier space as $\chi(\omega) \equiv h(\omega)/F(\omega)$, where $F$ is an external driving force. Adding this external force to the mode-dependent Langevin dynamics \eqref{eq:FourierLangevinEqDynamics}, we obtain in temporal Fourier space

\begin{equation}
-i\omega h_{\mathbf{q}}(\omega) = \Lambda_{q}\left[ - \left(\kappa q^4 +\sigma q^2 \right)h_{\mathbf{q}}(\omega) + F_{\mathbf{q}}(\omega) + \zeta^{\rm{th}}_{\mathbf{q}}(\omega)\right]
\end{equation}

Supposing that the driving force $F$ is much larger than thermal noise $\zeta^{\rm{th}}$, the mode-dependent response function is obtained as
\begin{equation}
\chi_{\mathbf{q}}(\omega) = \frac{\Lambda_{q}}{-i\omega + \omega_q}
\end{equation}
The response function can be separated into real and dissipative parts $\chi_{\mathbf{q}}(\omega) = \chi'_{\mathbf{q}}(\omega) + i\chi''_{\mathbf{q}}(\omega)$, leading to
\begin{equation}
\chi_{\mathbf{q}}'(\omega) = \frac{\omega_q\Lambda_{q}}{\omega^2 + \omega_{q}^2} \qquad\qquad
\chi_{\mathbf{q}}''(\omega) = \frac{\omega\Lambda_{q} }{\omega^2 + \omega_{q}^2}
\end{equation}

By identification with the auto-correlation function \eqref{eq:ModeDependent_CorrelationFunction}, and after summing on the modes $\mathbf{q}$, we deduce the fluctuation-dissipation relation
\begin{equation}
\label{eq:FluctuationDissipationRelation}
C(\omega) \equiv \left\langle \left| h(\omega)\right| ^2 \right\rangle = \frac{2k_{\rm{B}}T}{\omega}\chi''(\omega)
\end{equation}

\begin{figure}
	\centering
		\includegraphics[width=1.0\textwidth]{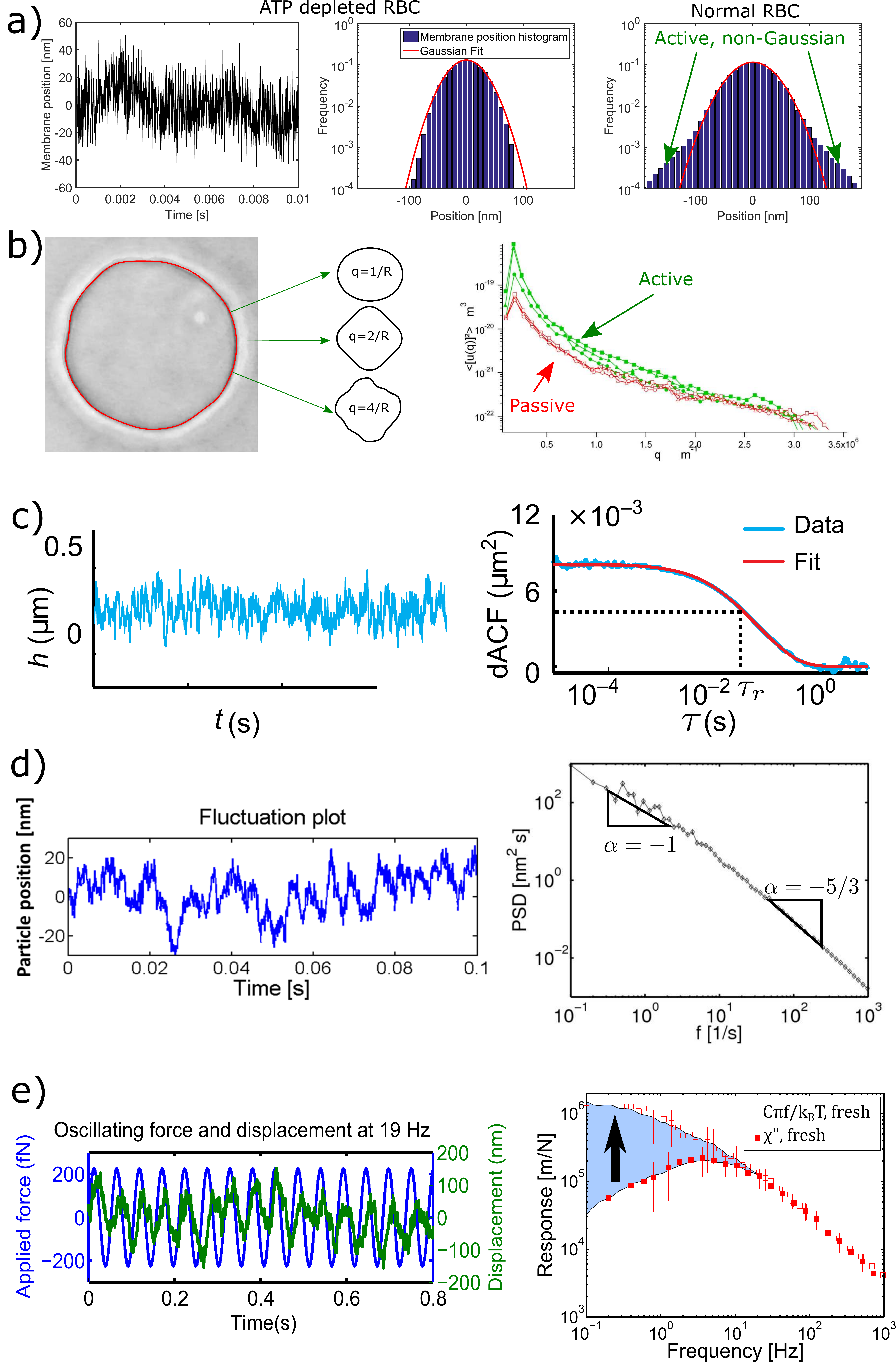}
	\caption{\textbf{Summary of the analysis methods to quantify active and passive fluctuations.} a) Using the histogram of the measured membrane fluctuations allows to quantify possible deviations from the expected Gaussian probability distribution. Commonly such non-Gaussian behavior is interpreted to result from active processes. b) The detailed mode analysis of the fluctuation in membranes shows an increased of fluctuation amplitude and its characteristic mode dependence (Data reproduced with permission from \cite{Faris:2009}) c) If only time dependent data is available, a common approach is to use the autocorrelation function and fit it with an exponential function, or more complex functions to describe the active component. (Data reproduced from \cite{Monzel:2015}) d) Another possible way to analyze time dependence membrane fluctuation is to determine the fluctuations spectrum. Membrane models can account for the different powerlaw observed in these measurements. e) A direct check of activity can be achieved by comparing the directly measured response functions (e.g. via optical tweezers) with the expected response function that is obtained using the free fluctuations spectrum and applying equilibrium physics. Possible differences as found in red blood cells directly show quantify the extend of activity. (Figures adapted from \cite{Turlier:2016})    }
	\label{fig:fig3}
\end{figure}

\subsection{Indirect fluctuation analysis}
In micropipette aspiration experiments, tension and excess areas constitute inherent average measures of the fluctuations over the full liposome area. As introduced in Figure \ref{fig:fig1} and in equation \eqref{eq:Surface_excess}, the membrane fluctuations lead to a difference between the projected membrane area $A$, and the total membrane area $\mathcal{A}$. Experimentally accessible is the change in excess area $\Delta \alpha \equiv \alpha_0 - \alpha$, obtained by measuring the radius of the liposome and the variation of the length of the membrane tongue $\delta L$ while aspirating with a micropipette. Here $\alpha_0$ is the excess area at the minimal tension $\sigma_0$ sufficient to suck up the liposome at the start of the experiment \cite{Manneville:2001}. As long as the membrane fluctuations remain in the entropic regime, we have shown above that the excess area takes the form $\alpha = \frac{k_{\rm{B}}T}{8\pi\kappa}\ln\left(\frac{\kappa\,q_{\rm{min}}^2}{\sigma}\right)$, from which we easily deduce the excess area change upon aspiration
\begin{equation}
\Delta \alpha =\frac{k_{\rm{B}}T}{8\pi\kappa}\ln \left(\frac{\sigma}{\sigma_0}\right)
\end{equation} 
The tension is controlled by the aspiration pressure and can be calculated using Laplace's law \cite{Manneville:2001}. In the experimental procedure the tension is systematically increased and the resulting excess area change is plotted over the tension which is normalized to the initial tension. This leads to datasets such as presented in Figure \ref{fig:fig2}a. In the case of active ion pumps in the membrane one finds systematically a reduced slope of the experimental curve \cite{Manneville:2001}. As the slope depends on the ratio between the temperature $T$ and the bending modulus $\kappa$ an initial analysis suggests investigation of these two parameters. To describe changes in the slope of the experimental data, an effective temperature $T_{\rm{eff}}$ is introduced. Since the physical reason for the excess area in passive systems is indeed a temperature dependent fluctuation, this intuitive modification provides a simple way to retain the analytical expressions. However, it should be mentioned that due to the spatial and temporal averaging any possible frequency dependence of $T_{\rm{eff}}$ is not accessible. This is important in the sense that the concept of temperature has no time dependence. Such an approach reflects the underlying assumption of statistical mechanics that thermal noise is a delta correlated stochastic variable with no inherent timescale. However, a dynamic analysis of the membrane fluctuations shows that the effective energy $E_{\rm{eff}}=k_{\rm{B}}T_{\rm{eff}}$ may be a frequency-dependent quantity that is hence qualitatively different from the classical concept of temperature. 
 
\subsection{Static analysis}
In contrast to the micropipette experiment, the static analysis of the data is often based on time-dependent raw data, such as video images. To obtain better statistics, time dependent data is condensed into histograms of the position or, in the extreme case, into a mean value and the standard deviation (Figure \ref{fig:fig3}a). This approach is very useful to reduce experimental noise if only a limited number of measurements are available. As the name points out, the static approach is fundamentally limited to static properties such as an elastic energy storage module that does not depend on timescales and that does not have inherent relaxation behavior. It should be mentioned that even such static measurements are still restricted to certain timescales that are given by the total acquisition time and the sampling rate of the raw data. Typically it assumes that the total acquisition time is large enough for the system to explore all possible conformations. This is for example the case when a probability distribution of visited membrane positions does not change shape for longer measurement times. In this view, an important point is that changes in friction or viscous properties lead to slower dynamics, which might require a critical check of recording timescales: at higher viscosity it can take much longer to explore all possible configurations.

Regarding image analysis, the membrane position is commonly detected by image processing and the amplitude for the different modes is extracted using Fourier analysis or spherical harmonics decomposition.  The average amplitude of the different modes is then plotted as a function of the mode number (Figure \ref{fig:fig3}b). Other possible analysis are cross-correlation curves, where distance-dependent correlation functions are calculated to understand the lateral length scale over which mechanical interaction are mediated by the membrane(Figure \ref{fig:fig3}c). In the case of active fluctuations initiated by bacteriorhodopsin, the activity enhances primarily lower modes (i.e. large wavelength) \cite{Faris:2009} (see Figure \ref{fig:fig3}b). This particular feature reflects in the time-domain, where low frequency fluctuations - corresponding to low modes - are also predominately enhanced by activity.  

Besides these direct observation of active fluctuations in model membranes, a static analysis was typically used to identify initial signs of active fluctuations in the red blood cell membrane. First measurements of relative membrane position showed relative changes in fluctuations upon ATP depletion, which was interpreted as a sign for activity. However, as the whole cell becomes stiffer upon depletion of intracellular ATP, a pure equilibrium interpretation may be proposed. Additional experimental results like changes of the static fluctuations amplitude when the buffer viscosity was changed did indeed hint for an active process \cite{Tuvia:1997}, however these experiments could not been reproduced. It is possible that insufficient recording time may not allow the membrane to explore all possible conformations in this pioneering experiment. 

More recently, the static analysis of membrane fluctuations was used to determine possible deviation from a Gaussian probability distribution of the membrane fluctuations. Such non-Gaussian contributions can be explained by active processes that generate membrane configurations hardly reachable from pure thermal agitation (Figure \ref{fig:fig3}a). A key feature of these non-Gaussian elements is that they are quite rare, which requires excellent statistics to be able to detect them. Using DRM a non-Gaussian parameter was indeed detected in the height fluctuations of red blood cells \cite{Park:2010}. In the case of ATP depletion, this non-Gaussian behavior was reduced, which was interpreted as a sign for activity. These results have been later confirmed by DODS. While this was an intriguing result, such a non-Gaussian behavior cannot be a conclusive proof of active fluctuations, as any nonlinear behavior in the membrane could lead to similar results. 

%In this view, the static fluctuation analysis is sufficient to provide quantitative measurements of activity in systems with known active contribution and where nonlinear effect are absent. A main advantage of a static analysis is that the obtained data is typically less affected by noise. 

\subsection{Dynamic analysis}
	\label{active}
When a large number of time-dependent measurements with good temporal resolution is available, it becomes reasonable to exploit time as additional dimension. Temporal analysis is interesting as it provides access to dynamic variables such as processes related to energy dissipation and friction. These parameters are fundamental to understand and model the mechanical processes involved in both passive and active fluctuations. This becomes evident when considering energy dissipation as the fundamental reason for thermal fluctuations. Briefly, thermal excitation of a membrane fluctuation implies an energy transfer from the thermal bath to the membrane which in turn requires a reduction of the bath's temperature. However, driving a movement by cooling a reservoir is in contradiction to the rules of thermodynamics. The fluctuation-dissipation theorem states that the same amount of energy is returned to the thermal bath by dissipating the energy stored in the movement, thus leading to a heating that exactly compensates the 'cooling' required to drive the fluctuations. Having this in mind, it becomes clear why the dissipation of active fluctuations gives a useful quantity to describe active processes, because it allows to measure the energy added to the system by the active process. Hence, to gain more information about the elastic but also dissipative properties of an active system, a time dependent analysis is vital. 

A common approach to look at time dependence is to study the autocorrelation function (ACF)  of the position $x(t)$, typically denoted $<x(\tau)x(0)>$. Depending on the type of analysis, the ACF is often normalized by subtracting the mean square of $x(t)$ and dividing by the variance of $x(t)$. When applying these normalizations the ACF becomes somewhat intuitive, as it varies only between $\pm1$, where a value of 1 marks a total correlation and -1 marks total anti-correlation. 
%At a time lag $\tau=0$ the normalized ACF is per definition 1, and typically decays rapidly for increasing lag times $\tau>0$. The decay of the ACF gives a measure of the mechanical memory in the system. Intuitively, it corresponds to an average similarity of the curve between time $t'$ and time $t'+\tau$. Typically, this influence decays with increasing $\tau$, which is visualized by the decay of the ACF. 
Often, an exponential decay of the ACF is found, which can be associated to a single relaxation process with timescale $\tau_r$ (Figure \ref{fig:fig3}c). In contrast, many systems show powerlaw behavior in the relaxation, which corresponds to a complex relaxation scheme with many processes. In the context of membranes, the ACF of a single mode typically relaxes exponentially. Every fluctuation mode has a characteristic wavelength $\lambda$ (typically reported: wavenumber $q=2\pi/\lambda$) that corresponds to a characteristic relaxation frequency $\omega_q=(\kappa q^3+\sigma q)/(4 \eta)$, as introduced in equation \eqref{eq:MembraneRelaxationRate}, where $\eta$ is the viscosity of the surrounding buffer solution. If the fluctuations are measured at one single position, they present the superposition of all accessible modes and hence the ACF results in a more complex powerlaw relaxation behavior. The overall advantage of the ACF lies in the visual and analytical interpretation of data containing a low number (1 to 3) of relaxation processes that are well separated in timescales, ideally by at least one order of magnitude. Also, the ACF can be efficiently calculated directly from the input data using dedicated hardware. It should be mentioned that the ACF function plays a key role in fluorescence correlation spectroscopy (FCS), which is a powerful tool in the study of lateral movements of lipids or membrane bound proteins. A key advantage of the correlation function is to combine temporal and spatial analysis by correlating two functions that represent the membrane movement at two different positions separated by a well defined distance. This two-point correlation can give information about timescales and distances over which mechanical forces can act on the membrane. Such analysis was used to give direct hints for active processes representing force dipoles with characteristic length scales \cite{Park:2010}. 

Besides the ACF, a second analysis type commonly used is the power spectral density (PSD), which is accessed by computing $C(\omega)=\frac{\tilde{x}(\omega)\times \tilde{x}(\omega)^*}{p \times s}$, where $\tilde{x}(\omega)$ is the Fourier transform of the time dependent signal $x(t)$, and $p$ and $s$ are the sample rate and number of datapoints respectively. The PSD refers to 'power', as it was originally used to get the electrical power of voltage measurements. Hence, strictly speaking the PSD of the membrane position is not a mechanical power, but reflects the fluctuation amplitude. Figure \ref{fig:fig3}d shows a typical PSD calculated from transverse fluctuations of a membrane. Important characteristic features are that the low frequencies provide a large amplitudes while at high frequencies the amplitudes are small. This is qualitatively explained by friction which prevents large amplitudes at high frequencies. The quantitative behavior of the PSD can be easily described in the case of an equilibrium quasi-flat membrane, as we presented in the paragraph \ref{sec:DynamicFluctuationSpectrum}: in the high-frequency regime, the spectrum is dominated by the bending modulus $\kappa$ and shows a $-5/3$ power-law, while in the low-frequency regime, dominated by the tension $\sigma$, the exponent changes to $-1$ (see equation \eqref{eq:PSDLimitScalingBehavior} and Figure \ref{fig:fig3}d). A quantitative model can be fit to the PSD to directly determine the mechanical properties of the membrane. In principle, the ACF and the PSD are exchangeable, as they are related to each other by a Fourier transformation in the time domain.  Advantages of the PSD are that single frequency noise sources, such as electronic 50 Hz noise show up as delta peaks and can be easily eliminated. The ACF however, allows to directly identify if a single or if multiple relaxation processes act on the system studied.  

In both cases, activity typically shows up as an increase in fluctuation amplitudes and possible difference in the power-laws from the equilibrium case. However, without a direct measure of the mechanical properties it is difficult to get a model-independent measure of active fluctuations.

\subsection{Direct measurement of the mechanical response function}
	\label{resp}
To experimentally prove the active nature of membrane fluctuations and to quantify the contribution of active forces to the fluctuations it is necessary to know the mechanical characteristics of the system. To measure mechanical properties, a well defined force is applied and its resulting deformation is measured. Since typical systems of interest such as cells and liposomes are marked by viscoelastic characteristics, it is important to measure time-dependent response functions. In principle this can be done by measuring the time dependent relaxation after application of a step force with an atomic force microscope, a magnetic tweezer, an optical tweezer or a calibrated micropipette. The time-resolution provided by step response is however generally not satisfactory in practice, and frequency response, where the system is probed successively for different frequencies, is generally preferred.
For biological membranes we are interested in the mechanical properties at the micrometer scale, and hence the corresponding methods are called microrheology. Classically microrheology is separated into active and passive methods. In active microrheology, an external force is applied to the system to measure its response, while in passive microrheology, the spontaneous fluctuations of the system are used to infer its mechanical properties. It should be mentioned here again that for active and in particular living systems, passive microrheology may not be used to infer mechanical response, unless the properties of the active process are known and integrated in the analysis.

In the context of active membranes, optical tweezers based microrheology has been used to measure the mechanical response of living cells, such as red blood cells and other eukaryotic cells \cite{Schlosser:2015}. In these measurements, the beads are attached to the plasma membrane and an oscillating force is applied while the response of the system is measured by following the bead movement. The advantage of the oscillatory driving is that noise and additional fluctuations can be removed by selecting the driving frequency $f = \omega/2\pi$ in the analysis. The response $\chi$ is then simply obtained by $\chi (f)=\tilde{x}(f)/F(f)$, where $F(f)$ is the oscillating driving force of frequency $f$. This experiment is repeated for driving forces of different oscillating period, to finally obtain the complex response function as a function of driving frequencies (Figure \ref{fig:fig3}e). This complex response function can be separated into its real and imaginary part, corresponding to the elastic ($\chi'(f)$) and dissipative ($\chi''(f)$) response. This approach was used to determine the dissipative response in the red blood cell membrane which was a key element to quantify active membrane fluctuations \cite{Turlier:2016}. Furthermore, the same approach allowed to determine the active fluctuations of granules in mouse oocytes cells \cite{Almonacid:2015} and of particles embedded in biomimetic actomyosin systems \cite{Mizuno:2007}.

\subsection{Test of fluctuation dissipation theorem}
	\label{FDT}
Combining the dynamic fluctuations analysis with mechanical response measurements allows to decouple the active part from the thermal driving in the membrane fluctuations. This presents a key element to evaluate the pertinence of theoretical models describing active fluctuations. Here we use the example of active fluctuations in the red blood cell membrane \cite{Turlier:2016}, but the measurement principle can be equally applied to other systems. In the case of red blood cells, four beads are attached to the membrane, and the red blood cell is held in space by trapping three handle beads. The free fluctuations are measured by decreasing the laser intensity on the fourth bead (probe bead) to a level where the optical detection system works reliably, but where the trapping force on the particle is negligible compared to thermal fluctuations. In this situation the movement of the bead can be assumed to be purely thermally driven, while being precisely recorded. The PSD $C(f)$ is calculated as described in section \ref{active}. In a second step of the experiment, the laser power on the probe bead is increased and the laser beam is moved in a sinusoidal fashion. The resulting oscillatory forces on the bead are recorded using the optical detection system. As described in section \ref{resp}, the frequency-dependent response function $\chi(f)$ is obtained. 

In an equilibrium situation, the free fluctuations and the dissipation are connected by the fluctuation dissipation theorem \eqref{eq:FluctuationDissipationRelation}, that we rewrite here as function of the frequency $f$: $C(f)=\frac{2 k_B T}{2\pi f} \chi''(f)$, where $\chi''(f)$ is the dissipative part of the response function. From an analysis point of view, the response function expected from an equilibrium system is obtained by measuring the PSD as $\chi''(f)= C(f) \pi f/k_{\rm{B}} T$. Plotting this function and the response function directly measured on the system provides two curves, which shall collapse if, and only if the system is at thermodynamic equilibrium (Figure \ref{fig:fig3}e). Any discrepancy between the two curves is therefore a clear evidence for an additional, non-equilibrium process driving active fluctuations. 

By analogy with the fluctuation-dissipation relation, one may define the effective energy and temperature as $E_{\rm{eff}} \equiv C(f) \pi f / \chi''(f) \equiv k_{\rm{B}}T_{\rm{eff}}$. It becomes clear that the effective temperature is, in general, timescale dependent, which is somewhat contradictory to the concept of temperature itself. By measuring $\chi''(f)$ and $C(f)$, the effective energy becomes directly experimentally accessible, which is an interesting method to investigate active fluctuations from a thermodynamic perspective, and to detect the frequency onset of active energy input in the system. For the red blood cell membrane, metabolic activity is found to only excite the membrane at short timescales: for frequencies higher than 10Hz, thermal fluctuations dominate the fluctuations, while at timescale lower than 100 ms, active contributions largely exceed thermal driving. Previous methods using the FDT, or passive microrheology, to infer the mechanical properties of red blood cell membrane have hence to be re-considered carefully.

\section{Theoretical models of active membrane}

The advent of theoretical modeling of active membranes may surely be associated to the seminal work of Prost and Bruinsma in 1996 \cite{Prost:1996}. In this landmark paper, an hydrodynamic model predicts the influence of ion channels gating activity on membrane fluctuations. Since then, a large set of models were proposed to account for direct or indirect sources of active noise in biological membranes and to analyze their specific features. All these hydrodynamic models take essentially a similar general form, where active and thermal noise are uncorrelated and may therefore be split into two distinct contributions. A common structure for active membrane models is therefore proposed. Then we discuss intrinsically active membrane models, where the active sources of noise comes from processes directly embedded in the bilayer, and extrinsically active membrane models, where activity is cytoskeleton-based but may be transmitted to the bilayer through mechanical coupling. 
Note that we do not cover here electrokinetic active membrane models, which have been recently reviewed elsewhere \cite{Lacoste:2014}. 
%We also mention only briefly numerical simulation models of active membranes in the perspectives.

	\subsection{General structure of active membrane models}
Starting from the equilibrium Langevin dynamics in Fourier space \eqref{eq:FourierLangevinEqDynamics},  the height dynamics for an active membrane may in general be cast into the generic following form 
\begin{equation}
\label{eq:ActiveMembraneLangevinEquation}
	\frac{\partial h_{\mathbf{q}}(t)}{\partial t} + \omega_{\mathbf{q}}h_{\mathbf{q}}(t) = \Lambda_{\mathbf{q}}\zeta_{\mathbf{q}}^{\rm{th}}(t) + \Lambda_{\mathbf{q}}\zeta_{\mathbf{q}}^{\rm{a}}(t)
\end{equation}
where $\omega_{\mathbf{q}}$ is the mechanical relaxation frequency of the membrane for the mode $\mathbf{q}$, $\zeta_{\mathbf{q}}^{\rm{th}}$ and $\zeta_{\mathbf{q}}^{\rm{a}}$ are thermal and active sources of noise, respectively, and $\Lambda_{\mathbf{q}}$ is a mode-dependent dissipative coefficient (e.g. the Oseen tensor for membranes surrounded by viscous fluids). In general, the membrane relaxation frequency takes the following form $\omega_{\mathbf{q}} = \Lambda_{\mathbf{q}}\,\delta \mathcal{H}/\delta h_{\mathbf{q}}(t)$, where $\mathcal{H}$ is the elastic energy of the membrane. 

The characteristics of the active noise source term $\zeta_{\mathbf{q}}^{\rm{a}}(t)$ are not constrained \textit{a priori} and may hence depend on the source of active forces, but as soon as this term is non-vanishing, one shall expect a violation of statistical equilibrium. We will hence describe in the following several models for the active source of noise in the membrane.

Note that equation \eqref{eq:ActiveMembraneLangevinEquation} can be readily generalized to quasi-spherical membranes with the use of spherical harmonics instead of Fourier decomposition \cite{Milner:1987, Lomholt:2006, Loubet:2012}.

	\subsection{Intrinsically active membrane models}
		\subsubsection{Ion channels shot noise activity}
		In the first active membrane model proposed by Prost and Bruinsma \cite{Prost:1996}, active fluctuations originate from the shot-noise activity of ion channels freely diffusing in the membrane (see Figure \ref{fig:fig1}b). 
		
		By switching stochastically between open and closed states, under the action of metabolic energy, these channels produce an additional source of noise, described by a two-state variable $S_k(t)=1$ if the ion channel $k$ is active and $0$ otherwise \footnote{Note that \textit{ion channels} do not require,  in general, metabolic energy consumption and their activity may hence be considered as passive. However, when a non-zero (electro)chemical potential difference is maintained across the membrane (generally through the action of ion pumps), their gating activity is expected to be of non-equilibrium character \cite{Gadsby:2009}.}. In agreement with single ion-channel gating measurements, the shot-noise is assumed to be exponentially correlated in time $g(t) = \left\langle S_k(t)S_k(0) \right\rangle - \left\langle S_k \right\rangle^2 = g(0)\,e^{-t/\tau_{\rm{a}}}$, with a typical correlation time $\tau_{\rm{a}}$. 
		
		By changing local osmolarity in the vicinity of the membrane, ion channels generate a local fluid pressure variation across the membrane of the form $\delta p(\mathbf{r},t) \propto f\sum_k S_k(t)\,\delta\big(\mathbf{r}-\mathbf{R}_k(t)\big)$, where $\mathbf{R}_k$ is the position of the ion channel $k$ in the membrane plane, and $f \sim k_{\rm{B}}T/w$ is the typical force amplitude exerted on the membrane of thickness $w$. Supposing the membrane semi-permeable with a permeability denoted $\lambda_{\rm{p}}$, the Langevin equation for the membrane height in Fourier space takes a similar form as \eqref{eq:ActiveMembraneLangevinEquation}
\begin{equation}
	\label{eq:Prost1996_LangevinDynamics}	
	\frac{\partial h_{\mathbf{q}}(t)}{\partial t} + \omega_{\mathbf{q}}h_{\mathbf{q}}(t) = \Lambda_q\zeta^{\rm{th}}_{\mathbf{q}}(t) + f\lambda_{\rm{p}}\sum_k S_k(t)\,e^{i\mathbf{q}\cdot\mathbf{R}_k(t)}
	\,\textrm{,}
\end{equation}
where $\omega_{\mathbf{q}} = \frac{\kappa}{4\eta}q^3 + \kappa\lambda_{\rm{p}}q^4$ is the sum of the relaxation frequency for an impermeable membrane with vanishing tension and an additional permeation term. The dissipative coefficient is the membrane permeability $\Lambda_{\mathbf{q}} = \lambda_{\rm{p}}$, and we can identify the active noise term as $\zeta_{\mathbf{q}}^{\rm{a}} (t) = f\sum_k S_k(t)e^{i\mathbf{q}\cdot\mathbf{R}_k(t)}$.

In the long wavelength limit, the model predicts that area density fluctuations of ion channels of active nature shall dominate the membrane fluctuation spectrum $\left\langle |h_{\mathbf{q}}|^2 \right\rangle \sim_{q\rightarrow 0} \frac{k_{\rm{B}}T}{\kappa}\left(q^{-4} + \xi^{-1}q^{-5}\right)$, where $\xi$ is a length scale proportional to the diffusion coefficient of channels in the membrane and inversely proportional to the shot noise correlation.

				This seminal work triggered a substantial theoretical interest for active membranes \cite{Turlier:2016, Ramaswamy:2000, Manneville:1999, Manneville:2001, Lomholt:2006, Loubet:2012, Chen:2004, Chen:2010, Gov:2004, Gov:2005, Gov:2006, Gov:2007, Lacoste:2005, Sankararaman:2002}, that we aim to briefly overview in the following.
				
		\subsubsection{Active curvature coupling}
\begin{figure}
	\centering
		\includegraphics[width=1.00\textwidth]{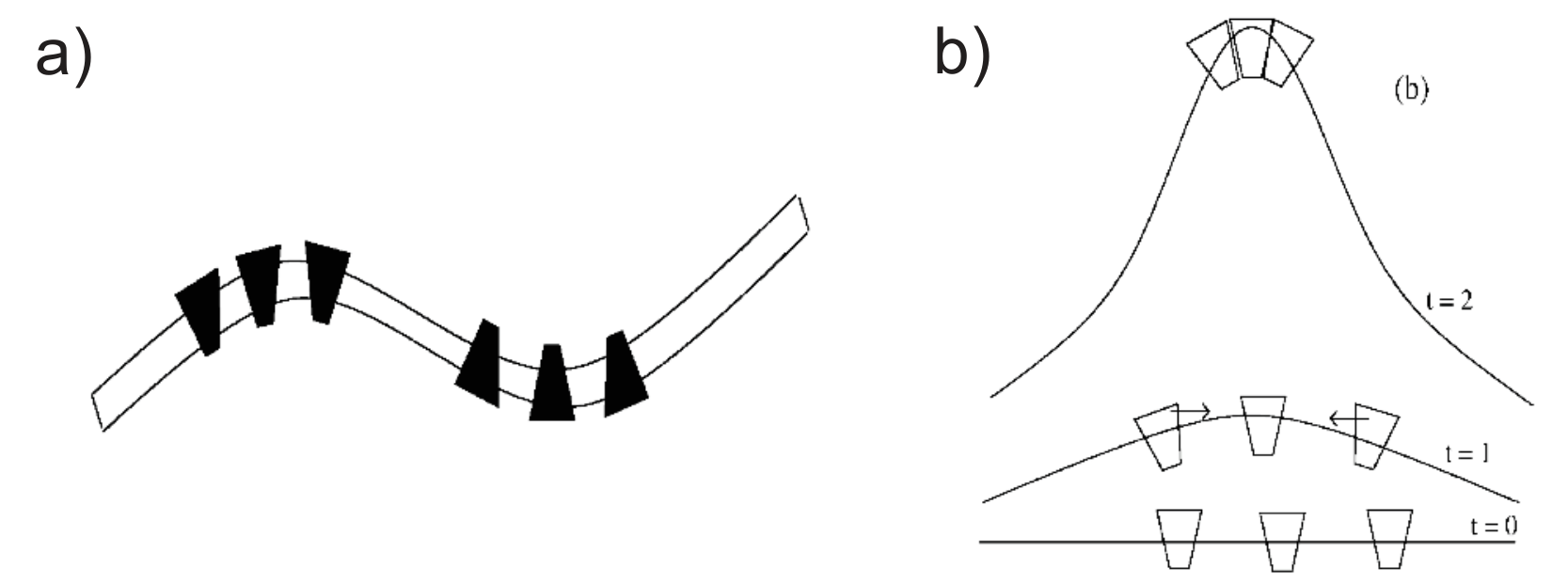}
	\caption{\textbf{Curvature coupling of ion pumps of asymmetric shape}. \textbf{a)} Asymmetric proteins are drawn to regions with curvature adapted to their shape. \textbf{b)} An instability may develop if the curvature produced by local pumping has the right sign to attract more pumps. (Figures reproduced from \cite{Ramaswamy:2000})}
	\label{fig:fig4}
\end{figure}

		The first experimental realization of active membranes \textit{in vitro} was done by reconstituting the transmembrane proton pump bacteriorhodopsin in giant unilamellar vesicles \cite{Manneville:1999, Manneville:2001}. A new theoretical model, considering the coupling between pumps activity and membrane curvature was conjointly proposed by Ramaswamy, Toner and Prost, to explain the experimental results \cite{Ramaswamy:2000}. The new ingredient added to the original model of Prost and Bruinsma \cite{Prost:1996} is an intrinsic asymmetry in the shape of ion pumps, inducing their preference either for positive or for negative membrane curvature regions (see Figure \ref{fig:fig4}a).
	
		A signed density of proteins $\psi(\mathbf{r},t) = n^{+}(\mathbf{r},t) - n^{-}(\mathbf{r},t)$, measuring the local difference between proteins with preference for positive curvature relative to proteins with preference to negative curvature, is introduced, and is coupled to the membrane elasticity in equation \eqref{eq:HelfrichHamiltonian_MongeRepresentation} up to second order in the variables
		\begin{equation}
		\mathcal{H}(h,\psi) = \frac{1}{2}\int_A d^2\mathbf{r}\left\lbrace \kappa\big[\boldsymbol{\nabla}^2{h}(\mathbf{r})\big]^2 + \sigma\big[\boldsymbol{\nabla}h(\mathbf{r})\big]^2 + \mu\psi^2(\mathbf{r}) - 2\Xi\psi(\mathbf{r})\boldsymbol{\nabla}^2 h(\mathbf{r}) \right\rbrace
		\end{equation}
where $\mu$ is the susceptibility for the imbalance between curvature positive and negative proteins and $\Xi$ is the curvature coupling coefficient. 
To close the problem, a conservation law for $\psi(\mathbf{r},t)$ is needed:
\begin{equation}
	\frac{\partial \psi(\mathbf{r},t)}{\partial t} \sim \Gamma\Delta\left(\frac{\delta\mathcal{H}}{\delta\psi}\right) + \boldsymbol{\nabla}\cdot \boldsymbol{\zeta}_{\psi}^{\rm{th}}
\end{equation}
where $\Gamma = D/\mu$ is a mobility coefficient, with $D$ is the diffusion coefficient of proteins in the membrane.

\begin{figure}
	\centering
		\includegraphics[width=0.85\textwidth]{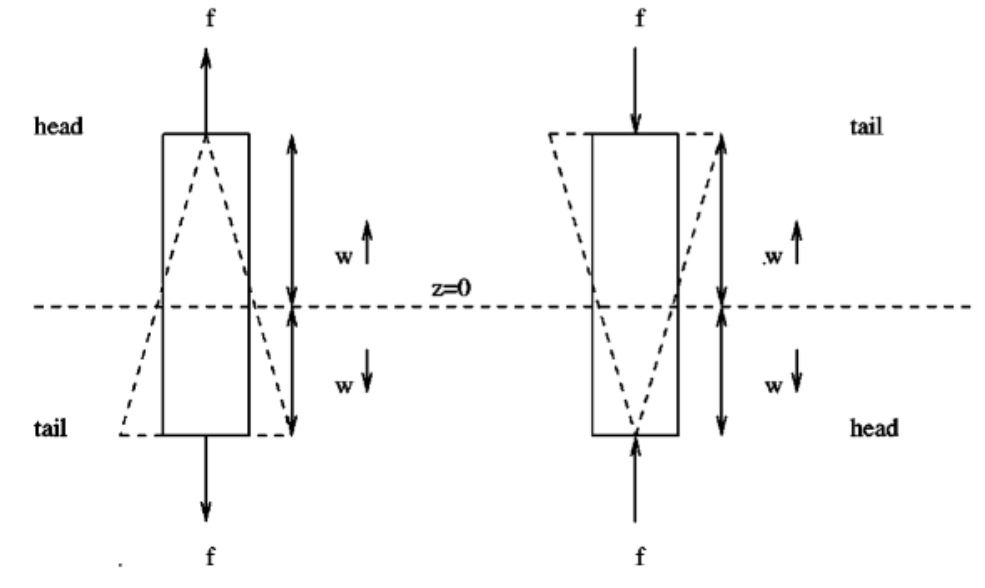}
	\caption{\textbf{Asymmetric dipole model used for active proteins:} the center of mass of the pumps is displaced above the bilayer midline and force center lies at distances $w^{\uparrow}$ and $w^{\downarrow}$ from this midline. The shape asymmetry of the pump is represented by asymmetric triangles. (Figure reproduced from \cite{Sankararaman:2002})}
	\label{fig:fig5}
\end{figure}

The activity of positive and negative pumps leads essentially to two additional forces normal to the membrane : 
\begin{itemize}
	\item An active permeation term of the form $\lambda_{\rm{p}}F_{\rm{a}}\psi(\mathbf{r},t)$ in Darcy's permeation equation across the membrane, where $F_{\rm{a}}$ is the elementary force transmitted to the membrane by the transfer of a proton.
	 \item An hydrodynamic active dipolar force density $F_a\big[ \delta(z-w^{\uparrow}) - \delta(z+w^{\downarrow})\big]\psi(\mathbf{r},t)$ in the force balance equation between the membrane and the surrounding fluid, where $w_{\uparrow}$ and $\, w_{\downarrow}$ are the distances of the center of mass of the ion pump relative to the membrane mid-plane, which are supposed distinct (see Figure \ref{fig:fig5}). 
\end{itemize} 
A numerical estimation of these two effects shows that the active permeation term may be omitted, in comparison to the active dipolar force, for the typical micron to submicron length scales relevant to biological membranes \cite{Manneville:2001}.

This set of equations results in two coupled Langevin dynamics for $h$ and $\psi$, which can be solved to obtain the membrane fluctuations autocorrelation and, eventually, an expression for the areal strain measured in micropipette experiments \cite{Manneville:2001}:
\begin{equation}
\Delta \alpha = \alpha_0 - \alpha  = \frac{k_{\rm{B}}T_{\rm{eff}}}{8\pi \kappa}\ln \frac{\sigma}{\sigma_0}
\end{equation}
The model predicts that the effect of pumps activity on the areal strain can be cast simply into an effective temperature $T_{\rm{eff}}$, in agreement with experimental results
\begin{equation}
\frac{T_{\rm{eff}}}{T} = \frac{\kappa}{\kappa"}\left( 1 + \mathcal{P}_{\rm{a}}\frac{\mathcal{P}_{\rm{a}}d^2 - \Xi d}{\kappa' d} \right) 
\end{equation}
where $\kappa'$ and $\kappa"$ are renormalized bending moduli and $\mathcal{P}_{\rm{a}} = F_a\frac{\left(w^{\uparrow}\right)^2-\left(w^{\downarrow}\right)^2}{2w}$ is the work of active dipoles. This formula shows that ion pumps will give rise to active fluctuations in the membrane only in the presence of an asymmetry $w^{\uparrow}-w^{\downarrow}\neq 0$ in the protein configuration within the bilayer.
	
The coupling between mobile force centers and membrane curvature leads furthermore to a new and rich physical behavior, exhibiting localized instabilities and traveling waves \cite{Ramaswamy:2000}. For example, ion pumps with a preference for positive curvature regions may trigger an instability: by increasing local membrane curvature through their pumping activity, they will attract more pumps and amplify the curvature even greater, as sketched on Figure \ref{fig:fig4}b. 

The model of Ramaswamy, Toner and Prost, does not consider, however, the random fluctuations in the pump activities (shot noise), introduced by Prost and Bruinsma \cite{Prost:1996}. In a subsequent paper, Lacoste and Lau addressed therefore simultaneously the effects of curvature couplings and pump activity fluctuations. While their model predicts similar local instability and travelling wave behaviors, it shows that shot noise effects are essential to consider for dynamical measurements of membrane fluctuations, as they lead to different scaling laws at short time scales \cite{Lacoste:2005}.

		\subsubsection{'Direct' vs. 'curvature force' and monopole vs. dipole}
If one neglects density fluctuation effects due to the diffusion of active proteins in the lipid bilayer, the active noise term can generally be assumed to be uncorrelated in space. In analogy with the shot-noise dynamics of ion channels, many models suppose active forces to be exponentially correlated in time, which is the property typically expected for a protein switching between "on" and "off" states \cite{Gardiner:1985}
\begin{equation}
\left\langle \zeta_{\mathbf{q}}^{\rm{a}}(t)\zeta_{\mathbf{q}'}^{\rm{a}}(t') \right\rangle = \Gamma^{\rm{a}}_{q}\delta_{\mathbf{q}+\mathbf{q}'}e^{-(t-t')/\tau_{\rm{a}}}
\label{eq:ActiveNoiseCorrelation}
\end{equation}
$\tau^{\rm{a}}$ is a typical active timescale, which may be defined from "on" and "off" rates $k_{\rm{on}}$ and $k_{\rm{off}}$ for a two-states metabolic process $\tau^{\rm{a}} \sim \big( k_{\rm{on}} + k_{\rm{off}}\big)^{-1}$, and $\Gamma^{\rm{a}}_q$ is the amplitude of the active noise, which may depend on the fluctuation mode $\mathbf{q}$.

Supposing that the elasticity of the membrane reduces to the Helfrich energy, the active membrane fluctuation spectrum can be calculated from equations \eqref{eq:EquilibriumNoiseCorrelation}, \eqref{eq:ActiveMembraneLangevinEquation} and \eqref{eq:ActiveNoiseCorrelation}:
\begin{equation}
\label{eq:ActiveFluctuationSpectrum}
\left\langle |h_q(\omega)|^2\right\rangle = \frac{2k_{\rm{B}}T\Lambda_q}{\omega^2 + \omega_q^2} + \frac{1}{\omega^2 + \omega_q^2}\frac{2\tau_{\rm{a}}\,\Lambda_q^2\Gamma_{q}^{\rm{a}}}{1+\omega^2\tau_{\rm{a}}^2}
\end{equation}
where the intrinsic membrane relaxation $\omega_{q}$ was defined in equation \eqref{eq:MembraneRelaxationRate}.

In \cite{Gov:2004}, Gov made a important distinction between two possible types of active forces, depending whether the  activity of proteins embedded in the membrane couples or not to the membrane curvature: 

\begin{itemize}
	
	\item For a so-called 'direct force', a random force of non-equilibrium origin is applied locally and directly to the membrane \cite{Gov:2004}. In this case, the active noise term $\zeta^{\rm{a}}_{\mathbf{q}}$ can be considered as an instantaneous "kick" on the membrane. The mean squared amplitude of the active noise is then independent of the mode $\mathbf{q}$ of membrane deformation and reads
	\begin{equation}
	\Gamma^{\rm{a}}_q = \rho_{\rm{a}}F_{\rm{a}}^2
	\label{eq:Direct_Active_Force}
	\end{equation}
	$F_a$ is the magnitude of the active force exerted on the membrane and $\rho_{\rm{a}} \sim N / A$ is the density of active proteins in the membrane.
	\\
	\item A 'curvature force' is on a contrary an active random contribution $c(\mathbf{r},t)$ to the spontaneous curvature, which may be introduced by generalizing the bending energy of the membrane as follows $\int_Ad\mathbf{r}\left\lbrace \frac{\kappa}{2}\left[\boldsymbol{\nabla}^2h(\mathbf{r}) - c(\mathbf{r},t) \right]^2\right\rbrace$ \cite{Lin:2006}.
	
	As a result, an additional random force density term appears in the right-hand side of the Langevin equation for the membrane \eqref{eq:ActiveMembraneLangevinEquation} in the form $\zeta^{\rm{a}}_{\mathbf{q}}(t) = - \kappa q^2 c_{\mathbf{q}}(t)$. Supposing that each active protein $i$ may switch metabolically between positive and negative spontaneous curvatures $+c_0$ and $-c_0$, we can write the time correlation function for the curvature of a single protein $i$ as $\left\langle c^i_0(t) c^i_0(t')\right\rangle = c_0^2 e^{-(t-t')/\tau_{\rm{a}}}$, where $\tau_{\rm{a}}$ is, like previously, a typical active timescale for the switching process. Summing over the active proteins in the membrane, the mean squared amplitude of the active noise now depends explicitly on the mode $q$
	\begin{equation}
	\Gamma^{\rm{a}}_q = \rho_{\rm{a}}\left(\kappa c_0 b^2q^2\right)^2
	\end{equation}	  
where $b^2$ is the typical surface area occupied by an active protein.

\end{itemize}
'Direct force' and 'curvature force' activities are hence predicted to produce different scaling behaviors for the membrane fluctuation spectrum in the limits of large and small wavelengths \cite{Gov:2004}. In subsequent works, the same author and colleagues generalized the model to \emph{diffusing} active proteins coupling to membrane curvature \cite{Lin:2006}, and characterized the fluctuations of a membrane where the proteins are considered to be nucleators of actin filaments \cite{Gov:2006}. Other authors considered a generalization where the membrane curvature may feedback onto the conformational transition kinetics of the active inclusions \cite{Chen:2010}. 
\\

An important distinction has to be made between monopolar, dipolar or quadrupolar active contributions. 'Direct forces' are typically force monopoles, which suppose essentially that an extrinsic agent, like the cytoskeleton, can push or pull locally on the membrane. Indeed, the spatial integral of the force over the system \{protein + membrane + solvent\} should vanish by force balance. If an active protein exerts a force on the membrane and solvent, the latter have to exert an opposite force of same magnitude on the protein. This implies that the force density field of the proteins has zero monopole moment, unless the system is actually not isolated because extrinsic agents enter the force balance. In general, the first contribution of active membrane proteins is a force dipole \cite{Lomholt:2006}, which can be idealized by two force centers of opposite sign but equal magnitude embedded in the membrane (see Figure \ref{fig:fig5}). A dipolar contribution is also expected from a permeation force \cite{Loubet:2012}. In the absence of dipolar contribution, higher multipole moments may however contribute to the active membrane fluctuations, such as for 'curvature forces', which can actually be considered as quadrupoles \cite{Lomholt:2006, Loubet:2012}.

		\subsubsection{Non-equilibrium fluctuations and excess area}
		\label{sec:excess_area}
			%We note that in \cite{Lacoste:2005,Chen:2010} the active membrane models are constructed based on the idea of switching between several internal states of the active proteins. The membrane fluctuations become therefore non-equilibrium fluctuations because the transitions between the various internal states of the proteins do not obey detailed balance.

In the models described so far, the description of active fluctuations ignores the interplay between excess area and surface tension.
As introduced by Seifert for passive membranes \cite{Seifert:1995}, membrane tension shall be formally regarded as a Lagrange multiplier for the conservation of total membrane area, and it is therefore intimately related to the excess area. The excess area is, by definition, a function of the fluctuation amplitude, which is itself a function of the tension. 
For a quasi-planar membrane, we can deduce from equation \eqref{eq:Surface_excess}
\begin{equation}
\alpha = \frac{1}{2A}\sum_{\mathbf{q}}q^2\left\langle \left|h_{\mathbf{q}}(\sigma)\right|^2 \right\rangle
\end{equation}

Setting the value of membrane excess area, the membrane tension is therefore controlled by the amplitude of fluctuations. The previous relation has, in general, to be inverted numerically, but analytic formula may been obtained from a pertubative approach for three different limit cases depending on the value of the dimensionless parameter $ k_{\rm{B}}T/\kappa\alpha$. 
%$\tau \gg 2 $, $\tau \ll 1/\ln(\bar{q}_{\rm{max}})$ and  $ 1/\ln(\bar{q}_{\rm{max}}) \ll \tau  \ll 2 $, where $\tau \equiv kT/\kappa\alpha$ and $\bar{q}_{\rm{max}} \equiv \sqrt{A}/a$, where a is a microscopic length of the order of lipid size \cite{Seifert:1995}. 

Generalizing this approach to active membranes, Loubet and colleagues show that, by increasing the fluctuations amplitude, the presence of activity in the membrane shall, in general, increase the bilayer surface tension \cite{Loubet:2012}. Distinguishing short and long membrane relaxation times $\omega_q^{-1}$ relative to a typical active time scale $\tau_a$, they derived analytical formula for the bilayer tension as function of the excess area of active membranes with either monopolar, dipolar or quadrupolar types of active forces.

It should be noted that the phenomenological finding that the fluctuations increase due to activity may also be interpreted as a reduction in tension. This can be illustrated by recalling that with the same pulling pressure more hidden membrane can be pulled out of a vesicle \cite{Faris:2009} when active forces enhance membrane fluctuation. This shows that the interpretation of the activity in the context of classical equilibrium approach may be ambiguous.

	\subsection{Cytoskeleton-based active membranes}
		\subsubsection{Renormalization of membrane properties by the cytoskeleton}
			The cell cytoskeleton being tightly coupled to the lipid bilayer, biological membranes are generally composite materials, where the mechanics is a combination of bilayer and cytoskeleton mechanics. A perturbative approach of the problem is to start with the Helfrich description of membranes, and to study how the presence of a cytoskeleton may renormalize endogenous properties, such as membrane tension or bending rigidity, or may create new types of mechanical response, such as resistance to shear or confinement of membrane deformations. Most of the work on the subject has been applied to the red blood cell membrane, where the cytoskeleton is, in comparison to other cell types, a more simple structure, made of a triangular network of extensible spectrin filaments, anchored to the bilayer at junction points via transmembrane protein complexes. Generalizations of these concepts may open new perspectives to characterize membrane fluctuations in cell types possessing an actomyosin cytoskeleton.
		
		\paragraph{\textbf{Membrane confinement by the cytoskeleton}}
			The spectrin cytoskeleton in red blood cells has been proposed by Gov and colleagues to confine the bilayer fluctuations \cite{Gov:2003}, an effect that can be rendered by the addition of an harmonic potential to the Helfrich energy \eqref{eq:HelfrichHamiltonian_MongeRepresentation}
\begin{equation}
			V = \frac{1}{2}\int_A d\mathbf{r}\,\gamma\, h(\mathbf{r})^2
\end{equation}
The confinement term, of amplitude $\gamma$, constrains the mean squared amplitude of fluctuations to be equal to $d^2 = \frac{1}{8}k_{\rm{B}}T/\sqrt{\gamma\kappa}$. This is equivalent to consider that the spectrin cytoskeleton acts as a rigid plane at $h=0$ that maintains the membrane at an average distance $d$ via an harmonic restoring force. The static fluctuation spectrum of the confined membrane in Fourier space can be calculated at equilibrium as
\begin{equation}
\left\langle \left| h_{\mathbf{q}}\right|^2 \right\rangle = \frac{k_{\rm{B}}T}{\gamma + \sigma q^2 + \kappa q^4}
\end{equation}

\begin{figure}
	\centering
		\includegraphics[width=1.0\textwidth]{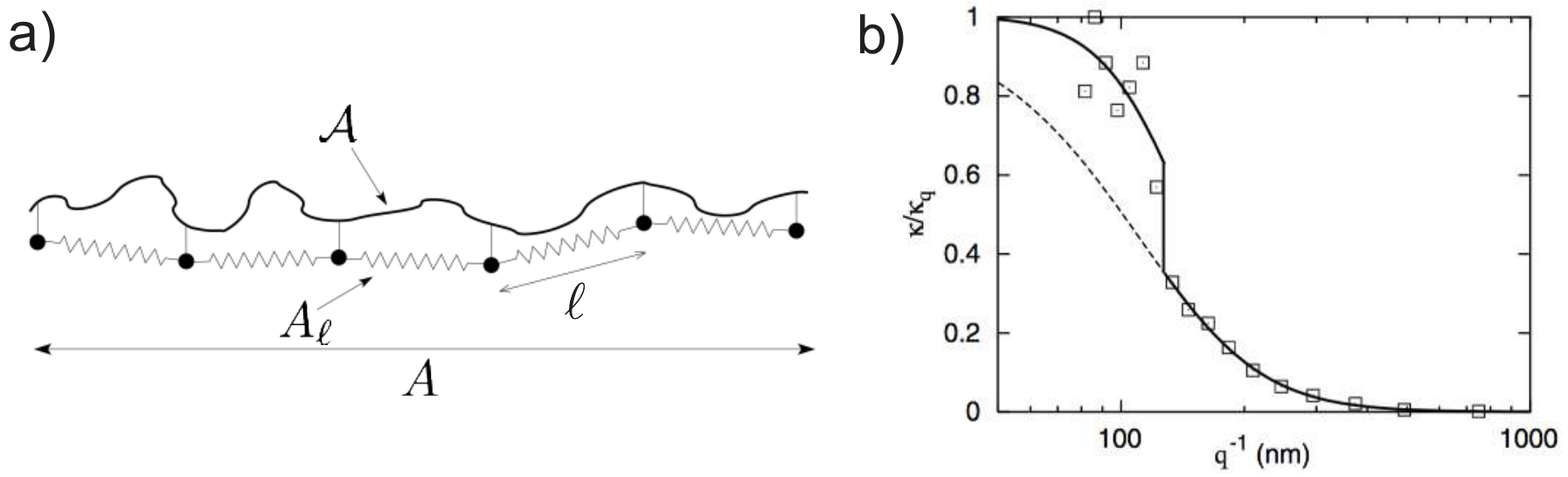}
	\caption{\textbf{Composite structure and static fluctuation spectrum of a quasi-planar red blood cell membrane}. \textbf{a)} Sketch of a nearly planar red blood cell membrane of total area $\mathcal{A}$, of coarse-grained area $A_{\ell}$ and projected area $A$. The spectrin filaments are represented by linear springs of length $\ell$ anchored in the bilayer. \textbf{b)} Fit (plain line) of the static fluctuation spectrum of a human red blood cell measured by \cite{Zilker:1987} with the model of Fournier et al. \cite{Fournier:2004} predicting a tension jump at $q^{-1}\approx 125 \rm{nm}$ due to the spectrin cytoskeleton elasticity. The dashed line extrapolates the large wavelengths fit, without tension jump. (Figures adapted or reproduced from \cite{Fournier:2004})}
	\label{fig:fig6}
\end{figure}

The confinement parameter $\gamma$ defines a new characteristic length $\lambda_{\gamma} = \left(\kappa/\gamma\right)^{1/4}$, which determines the wavelength onset for membrane confinement. In their study, Gov et al. show that adding a confinement parameter of the order of $\gamma \sim 10^{-7} \rm{J.m}^{-4}$ allows for a better fit of the mode-dependent experimental fluctuation spectrum measured by Zilker et al. in red blood cells \cite{Zilker:1987}, which displays an abrupt drop in fluctuations for wavelengths above $100\,\rm{nm}$ (see Figure \ref{fig:fig6}).

It should be noted that such confinement potential has been introduced originally in the context on membrane adhesion to a surface \cite{Radler:1995}, to render the combined effect of electrostatic attraction and steric repulsion (by long glycocalyx chains) of membrane to the surface \cite{Sackmann:2014}.

		\paragraph{\textbf{Effective membrane mechanics}}

A subsequent model was proposed by Fournier et al. to explain this jump in the fluctuation spectrum \cite{Fournier:2004}. In this model, the composite red blood cell membrane energy is supposed to be the sum of the Helfrich energy \eqref{eq:HelfrichHamiltonian_MongeRepresentation} and an elastic contribution from the spectrin cytoskeleton $\mathcal{H}_{\rm{el}} = \frac{1}{2}N k (\ell - \ell_0)^2$, supposed to be a perfect network of $N$ entropic springs of stiffness $k$ and of actual and resting lengths respectively $\ell$ and $\ell_0$. The authors show that this additional term leads to a jump in the membrane tension at wavelengths larger than the mesh size $\ell$
\begin{equation}
\Delta \sigma \sim \frac{1}{2}gk\left(1-\frac{\ell_0}{\ell}\right)
\end{equation}
where $g$ characterizes the topology of the network.

This model supposes fundamentally that the spectrin network is prestressed ($\ell \neq \ell_0$) and the best fit of the experimental data from Zilker et al. \cite{Zilker:1987} is obtained for a spectrin network under extension $\Delta \sigma = \sigma^{<} - \sigma^{>} \sim 1.6\times 10^{-6}\,\rm{N.m}^{-1}$ and for a lipid bilayer under compression $\sigma^{>} = \sigma \sim -0.8\times10^{-7}\,\rm{N.m}^{-1}$. The authors suggest that the spectrin cytoskeleton may indeed lead to extra-folding of the bilayer, therefore regulating the membrane tension both directly at large scales compared to the mesh size, and indirectly at shorter scales, via the membrane area constraint. In a following paper, a more detailed calculation at higher orders predicts that the bending modulus is also renormalized by the presence of the spectrin cytoskeleton \cite{Dubus:2006}.

		\subsubsection{Models of active red blood cell membrane fluctuations}

\begin{figure}[h!]
    \centering{\includegraphics[width=0.95\textwidth]{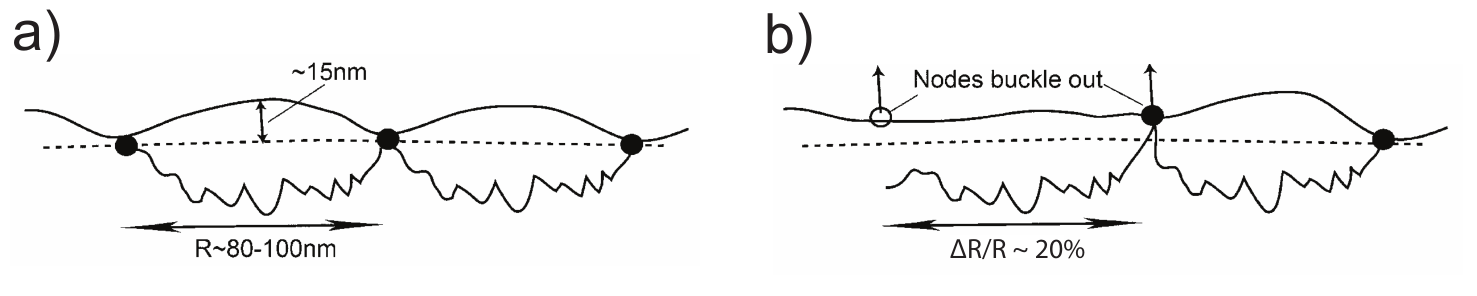}}
	\caption{\textbf{Active spectrin dissociation model for the red blood cell membrane.} \textbf{a)} Schematic side view of the red blood cell composite membrane with fully connected spectrin filaments. \textbf{b)} Sketch of the dissociation of a filament, which would generate a direct normal force on the bilayer. (Figures reproduced from \cite{Gov:2005}) }
	\label{fig:fig7}
\end{figure}	

In successive papers \cite{Gov:2004, Gov:2005, Gov:2007}, Gov \& Safran proposed a first model for active fluctuations in the red blood cell membrane, where the activity is supposed to originate from the spectrin cytoskeleton. Direct active forces ("kicks") on the membrane (see equation \eqref{eq:Direct_Active_Force}) are suggested to be triggered by the detachment, under ATP hydrolysis, of spectrin filament ends from the lipid bilayer. In average, filament detachment is also predicted to soften the red blood cell membrane by decreasing the stiffness of the spectrin cytoskeleton, which is assumed to be naturally prestretched. Balancing the tension developed by a stretched filament with the energy needed to buckle the membrane, the authors predict a steady-state prestretch of the spectrin network of approximately $20\%$ (see \cite{Gov:2005}). Assuming that detachment events are exponentially correlated in time, the non-thermal fluctuation spectrum originating from active direct forces are calculated in the same form as in equation \eqref{eq:ActiveFluctuationSpectrum}. In a subsequent work, Auth, Safran \& Gov calculated the entropic pressure exerted on the lipid bilayer by fluctuations of the spectrin filaments themselves \cite{Auth:2007_NJP} and they derived the fluctuation spectrum of coupled solid and fluid membranes maintained at fixed distance. They showed that this composite system may be described as a single polymerized membrane with renormalized bending rigidity \cite{Auth:2007_PRE}.

The model of Gov \& Safran assumes that active detachment of spectrin filaments may drive direct normal forces onto the membrane, but the precise microscopic mechanism of momentum transfer is not explicitly derived. In an alternative approach \cite{Turlier:2016}, Turlier et al. recently proposed a new active model for the composite red blood cell membrane, where the mechanical coupling between the lipid bilayer and the spectrin cytoskeleton is precisely derived, and the specrin activity is described in a more generic manner. 

\begin{figure}[h!]
    \centering{\includegraphics[width=1.02\textwidth]{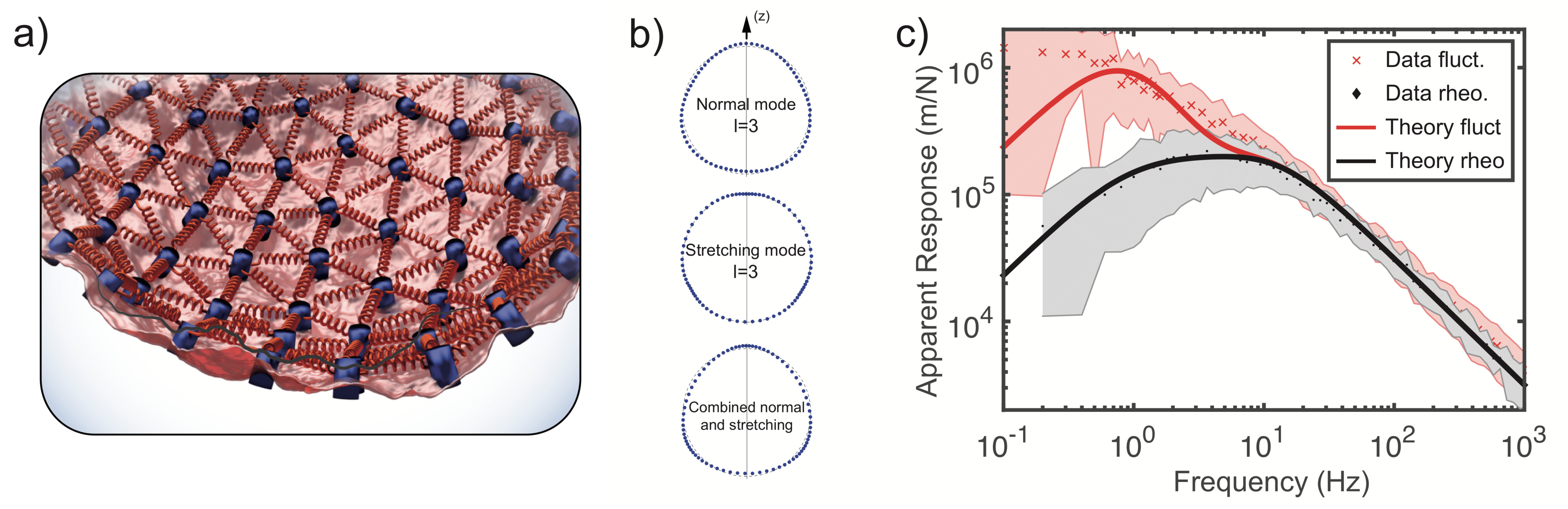}}
	\caption{\textbf{Active quasi-spherical model of the composite red blood cell membrane} \textbf{a)} Schematic representation of the red blood cell membrane composed of a lipid bilayer and a regular triangular network of spring-like spectrin filaments. \textbf{b)} The composite membrane deformation can be separated in bending and stretching modes, illustrated for the spherical harmonic $l = 3$. \textbf{c)} The analytic model (solid lines) can reproduce the experimental fluctuations and response data (crosses).  (Figures adapted from \cite{Turlier:2016}) }
	\label{fig:fig8}
\end{figure}	

Since all metabolic events identified in the spectrin network, or in its anchoring proteins, have been associated with a decreased mechanical strength of the membrane it is supposed that any phosphorylation shall lead to a local decrease of the network shear modulus, the single parameter necessary to characterize the spectrin network mechanics. A simple two-state dynamics is assumed for simplicity, and is characterized by two transition rates $k_{\rm{on}}$ and $k_{\rm{off}}$, defining the active timescale $\tau_{\rm{a}} = \left( k_{\rm{on}} + k_{\rm{off}} \right)^{-1}$ and the fraction of active sites $\left\langle n_{\rm{a}}\right\rangle = k_{\rm{on}}/\left( k_{\rm{on}} + k_{\rm{off}} \right)$. The network shear modulus is then supposed to fluctuate around a mean value that decreases with metabolic activity $\left\langle \mu \right\rangle = \mu_0\left(1 - \left\langle n_{\rm{a}}\right\rangle\right)$.
In line with previous studies, the spectrin cytoskeleton is idealized as a perfect triangular network of linear, prestressed springs (see Figure \ref{fig:fig8}a). As the prestress is supposed to be isotropic, the discrete elastic energy of the network can be homogenized into a continuous isotropic elastic membrane. Its energy is expressed as a function of the incremental deformation from the prestressed state and takes an Hookean form with an additional prestress term. 
%Integrated over the membrane surface, it reads $\mathcal{H}_{\rm{el}} = \int_A \epsilon_i + Se_{ii} + M\left(1/2(e_{ij})^2 + e^2_{ij}\right)$, where $\epsilon_i$ is a constant energy density due to prestress and $e_{ij}$ is the nonlinear incremental membrane strain tensor. 

The mechanics of the prestressed network is characterized by an effective spectrin tension S and an incremental shear modulus M, which are functions of the original shear modulus $\mu_0$, of the activity $\left\langle n_{\rm{a}} \right\rangle$ and of a prestretch ratio. The membrane is supposed quasi-spherical and due to curvature, stretching and bending modes of deformations of the network are linearly coupled, as illustrated on Figure \ref{fig:fig8}b for the mode $l = 3$. Since the lipid bilayer is tangentially fluid, the proteins anchoring the network to the bilayer may have non-zero sliding velocity relative to lipids, which, in turn, exert a drag force on the network. This drag force is found to be the dominant dissipative contribution slowing down membrane stretching at large wavelengths. On the contrary, normal deformations are balanced by viscous forces from surrounding fluids. In this context, one has to calculate explicitly the lateral lipid pressure - or instantaneous surface tension -, which acts as a Lagrange multiplier for local bilayer incompressibility (to distinguish from the global surface tension, which acts as a Lagrange multiplier for the total bilayer area). It turns out, from the calculation, that the lateral lipid pressure cancels systematically direct normal forces that may originate from active fluctuations in the spectrin mechanics. Since fluctuations in the network shear modulus lead to active forces in both tangential and normal directions, an indirect source of active noise is however conserved in the normal direction thanks to the coupling between bending and stretching modes via the curvature. 
Decomposing the deformations in spherical harmonics modes $(l,\,m)$, the membrane shape fluctuations appear classically as a sum of thermal and active contributions, expressed here with the membrane response function and the spectrin metabolic activity, respectively:

\begin{equation}
 C_{lm}(f) = \frac{2k_{\rm{B}}T}{2\pi f}\chi_{lm}''(f) + \frac{2\left\langle n_{\rm{a}} \right\rangle\left( 1- \left\langle n_{\rm{a}} \right\rangle \right)\tau_{\rm{a}}}{1 + \left(2\pi f \tau_{a}\right)} \left| N_{lm}(f)^2 \right|^2
 \label{eq: }
\end{equation}
where $f = \omega/ 2\pi$ is the frequency, and $N_{lm}(f)$ captures the complex mode- and frequency-dependent propagation of tangential active noise into membrane shape fluctuations.

The model predicts that spectrin-based active fluctuations should vanish for quasi-planar geometries, and it anticipates higher fluctuations in more curved regions of the red blood cell membrane, in agreement with recent spatial interferometric measurements \cite{Park:2010}. It also shows clearly that a prestress in the network is the necessary ingredient for the emergence of spectrin-based active fluctuations, in agreement with previous hypotheses \cite{Gov:2005,Gov:2007}. It finally predicts that the network prestress may be maintained internally by an excess area of bilayer membrane, as suggested earlier by Fournier et al. \cite{Fournier:2004}. Inverting the relation between excess area and bilayer tension (see section \ref{sec:excess_area}), a negative bilayer tension of the order of $10^{-7}\,\rm{N.m}^{-1}$ is found by fitting experimental data. This analytical model reproduces fairly both the response function and active membrane fluctuations measured in the red blood cell membrane \cite{Turlier:2016}, as shown on Figure \ref{fig:fig8}c.

\section{Perspectives}
 
		\subsection{A physiological role for active membrane fluctuations?}

\begin{figure}[h]
    \centering{\includegraphics[width=1.02\textwidth]{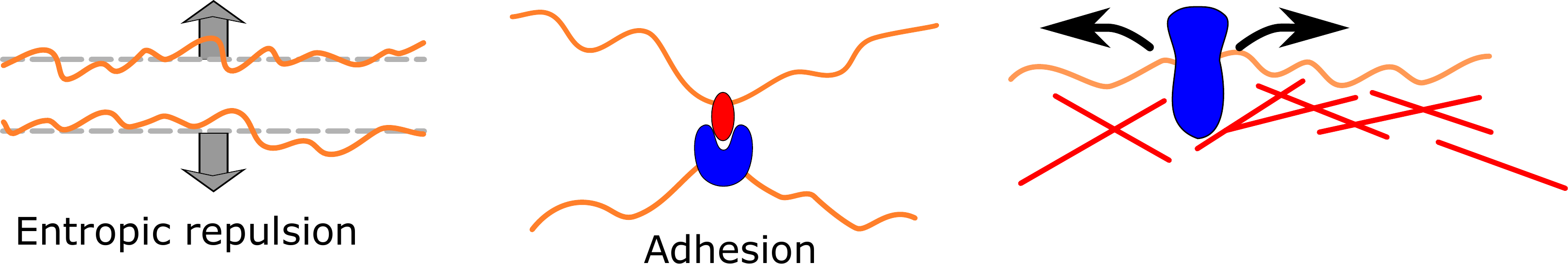}}
	\caption{\textbf{Possible biological functions of enhanced fluctuations related to membrane adhesion, repulsion and mixing.} The fluctuations do create an effective potential barrier that can prevent uncontrolled membrane fusion. On the contrary, the large amplitudes of fluctuation may allow receptors and trans-membrane proteins to explore larger volumes, and thus facilitate the controlled binding and adhesion between membranes. Additionally, the active forces might increase lipid mixing and help protein transport in the membrane again the cortical cytoskeleton (red lines).    }
	\label{fig:fig9}
\end{figure}

An interesting and still overlooked question is the potential physiological role of active membrane fluctuations. The occurrence of active fluctuations does not substantiate by itself any functional role in biology, and active noise may be simply a unavoidable byproduct of normal active processes in the cell. Yet, as noise is ubiquitous at this scale, cells may also control and take advantage of active noise, to actively facilitate or regulate other essential processes. It remains difficult, both experimentally and theoretically, to discriminate the potential role played by active fluctuations, from the main purpose of the active process considered. While still unproven, a number of possible physiological roles of membrane flickering have been suggested. As fluctuations increase membrane movement and fluidity, it should help mixing lipids and proteins within the membrane, an important property for cellular homeostasis (Figure \ref{fig:fig9})\cite{Lin:2004}. Additionally, the fluctuations might help larger membrane bound proteins to be transported laterally as it can help to overcome possible steric obstacles such as intracellular cytoskeletal components that might prevent or reduce lateral transport \cite{Lin:2004}. A further possible role lies in the interaction with other, intracellular as well as extracellular membranes. Depending on the function, the fluctuations may help surface bound receptors to interact with other membranes, as the fluctuations allow to explore a larger region (Figure \ref{fig:fig9}). On the other hand, active fluctuations may help to suppress nonspecific interaction by creating an effective repulsive force when an obstacle enters the region of the fluctuations \cite{Evans:1986,Prost:1998}. Finally, it was observed that active fluctuations modify the effective tension that is measured on membranes \cite{Betz:2009}, while on the other hand, membrane tension is known to be an important mechanical parameter for a number of cellular functions ranging from cell motility to endo/exocytosis and mechanosensing.

		\subsection{Active fluctuations in membrane adhesion processes}
		
Membrane adhesion is the physical process of interaction and attachment of a membrane to a surface, substrate or another membrane (Figure \ref{fig:fig9}). In cells, membrane adhesion processes play essential roles in cell migration, cell-cell interaction or bilayer-cytoskeleton coupling and are therefore tightly regulated by various membrane proteins (cadherins, integrins, ERM proteins) \cite{Sackmann:2014}. Membrane adhesion is, in general, controlled by a competition between attractive forces at short range (stickers), repulsive forces at intermediate distances (spacers) and elastic stresses coming from membrane deformation \cite{Bell:1988, Evans:1985}.
%In biological membranes we can distinguish two kinds of membrane-anchored molecules, stickers and repellers. Sticker molecules mediate attractive interactions between the membrane and the adjacent substrate, while repellers act as repulsive spacers. 
Depending on their amplitude, membrane fluctuations may play antagonist roles towards adhesion: moderate fluctuations may assist the nucleation \cite{Bihr:2012} and the expansion of adhesion domains, but at higher amplitudes fluctuations will compete with attractive forces and promote detachment of the membrane. The control of active fluctuations becomes in this context essential, but the role of activity in membrane adhesion has remained widely overlooked so far. Most of the theoretical models for membrane adhesion have considered membranes at equilibrium, which is justified when the focus is on biomimetic systems \cite{Bruinsma:2000, Bihr:2012}. To generalize these concepts to biological membranes, it will become critical to evaluate experimentally and theoretically the influence of active fluctuations on adhesion processes \cite{Weikl:2009}.

		\subsection{Fluctuations of membranes with an actomyosin cytoskeleton}

Most animal cell types possess a cortex of actomyosin, thin layer of actin filaments with embedded molecular motors, which lies beneath the membrane. The actomyosin cortex is tightly connected to the lipid bilayer via different proteins, the principal family being formed by the ERM (Ezrin-Radixin-Moesin) \cite{Fehon:2010}. Regulated by phosphorylation, the connection between ERM and the cortex is dynamic and non-equilibrium by nature, providing a first possible source of active forces in the membrane. The active regulation of membrane-cortex adhesion is particularly important for the formation of blebs, membrane bulges originating from local cortex detachment \cite{Charras:2008,Peukes:2014,Alert:2016}. But the cortex itself is notoriously very dynamic, with several processes requiring metabolic energy consumption, such as actin polymerization and myosin motor activity. All these processes constitute potential sources of active membrane fluctuations, covering a large spectrum of length and timescales. To disentangle and characterize precisely the various sources of activity in composite bilayer-cortex membranes, a substantial mutual effort from biologists, physicists and theorists will be required over the next years, and precise numerical modeling will become critical \cite{Turlier:2016,Fedosov:2010}.

%		\subsection{Numerical models of active membranes}
%Zhang/Brown 2008 Langevin dynamics \cite{Zhang:2008}
%Fedosov, DPD simulations \cite{Fedosov:2010, Turlier:2016}

\begin{acknowledgement}
H. Turlier acknowledges support from the CNRS/Inserm program ATIP-Avenir, from the Bettencourt-Schueller Foundation and from the Coll\`{e}ge de France.
T. Betz is supported by the Deutsche Forschungsgemeinschaft (DFG), Cells-in-Motion Cluster of Excellence (EXC 1003-CiM), University of M\"{u}nster, Germany. 
\end{acknowledgement}
%

%Check for the refs
%\input{referenc}

%Check for the refs

\end{document}